\documentclass[useAMS,usenatbib]{mn2e}
\usepackage{graphicx}
\usepackage{times}
\usepackage[dvips]{color}

\def\mmm{$(m-M)_0$}
\def\ebv{$E(B-V)$~}

\def\vi{$V-I$}
\def\bv{$B-V$}

\def\msun{M$_{\odot}$}
\def\gsim{\;\lower.6ex\hbox{$\sim$}\kern-7.75pt\raise.65ex\hbox{$>$}\;}
\def\lsim{\;\lower.6ex\hbox{$\sim$}\kern-7.75pt\raise.65ex\hbox{$<$}\;}

\title[To 2, Be 20, and  Be 66]{
Old open clusters and the Galactic metallicity gradient:  
Berkeley 20, Berkeley 66, and Tombaugh 2\thanks{ 
Based on observations made with the Italian
Telescopio Nazionale Galileo (TNG) operated on the island of La Palma by
the Fundaci\'on Galileo Galilei of the INAF (Istituto Nazionale di
Astrofisica) at the Spanish Observatorio del Roque de los Muchachos of
the Instituto de Astrofisica de Canarias, and on observations obtained at the
ESO telescopes in La Silla (Chile) under programmes 67.D-0014, 68.D-0222}
}

\author[Andreuzzi et al.]{
  Gloria Andreuzzi$^{1,2}$\thanks{E-mail: 
  andreuzzi@tng.iac.es (GA), 
  angela.bragaglia@oabo.inaf.it (AB), 
  monica.tosi@oabo.inaf.it (MT)
  gmarconi@eso.org (GM)}, 
 Angela Bragaglia$^{3}$,  Monica Tosi$^{3}$
 and Gianni Marconi$^{4}$ \\
 \\
 $^1$  Fundaci\'on Galileo Galilei - INAF, 
 Rambla Jos\'e Ana Fern\'andez P\'erez, 7, 38712 Bre\~na Baja, TF (Spain)\\
 $^2$ INAF--Osservatorio Astronomico di Roma, Via dell'Osservatorio 5, 
      I-00040 Monte Porzio (Italy) \\
 $^{3}$ INAF--Osservatorio Astronomico di Bologna, Via Ranzani 1, I-40127 Bologna
      (Italy) \\
  $^4$ ESO, Alonso de Cordova 3107, Vitacura, Santiago, Chile \\
     }

\date{}

\begin{document}
\maketitle

\begin{abstract}
To study the crucial range of Galactocentric distances between 12 and 16 kpc, where 
little information is available, 
we have obtained $VI$ CCD imaging of Berkeley~20 and $BVI$ CCD imaging
of Berkeley~66 and Tombaugh~2, three distant, old open clusters.
Using the synthetic colour magnitude diagram (CMD) technique with three types 
of evolutionary tracks of  different metallicities, we have determined age, 
distance, reddening and indicative metallicity of these systems. 
The CMD of Be~20 is best reproduced by stellar models with a metallicity about 
half of solar (Z=0.008 or 0.01), in perfect agreement with high resolution 
spectroscopic estimates. Its age $\tau$ is between 5 and 6 Gyr from stellar models 
with overshooting and between 4.3 and 4.5 Gyr from models without it.
The distance modulus from the best fitting models is
always (m-M)$_0$=14.7 (corresponding to a Galactocentric radius of about 16
kpc), and the reddening \ebv ranges between 0.13 and 0.16.
A slightly lower metallicity (Z$\simeq$0.006) appears to be more appropriate 
for Be~66. This cluster is younger, $\tau$=3 Gyr, and closer, (m-M)$_0$=13.3
(i.e., at 12 kpc from the Galactic centre), than Be~20, and suffers from high
extinction, 1.2$\leq$ \ebv $\leq$1.3, variable at the 2-3 per cent level. 
Finally, the results for To~2 indicate that it is an intermediate age cluster, with $\tau$ about 1.4 Gyr or
1.6-1.8 Gyr for models without and with overshooting, respectively. The metallicity is about 
half of solar (Z=0.006 to 0.01), in agreement with spectroscopic determinations. The distance modulus is  (m-M)$_0$=14.5, implying a distance of about 14 kpc from the Galactic centre; the reddening \ebv
is 0.31-0.4, depending on the model and metallicity, with a preferred value around 0.34.

\end{abstract}


\section{Introduction} \label{intro}

Open clusters (OCs) are very good tracers of the Galactic disc properties, of its formation and evolution
\citep[e.g.][]{panagiatosi,friel95,taat,fbh}. In particular, OCs can be used to study the metallicity distribution 
in the disc and its possible evolution with time.
With the BOCCE (Bologna Open Cluster Chemical Evolution) project, described in detail by \cite{bt06}, we are
deriving precise and homogeneous ages, distances, reddenings and chemical
abundances for a large sample of OCs. 
The final goal is to study the present status of the Galactic disc, its 
formation and evolution. Since the least known epochs of the disc evolution
are the earliest ones, we put particular attention to the study of old clusters. 
Adding the present three old systems to those already described
in \cite{bt06}, to Be~17 \citep*{bra06a}, Be~32 and King~11 \citep*{tosi07},
we have already examined 17 clusters older than 1 Gyr, out of the about 190
listed in the  \cite{dias02} catalogue. 
 
As part of this project, we present here a photometric study of  three clusters:
(a) Berkeley~20 ($l=203.^\circ48$, $b=-17.^\circ37$, in the third Galactic quadrant); 
(b) Berkeley~66 ($l=139.^\circ43$, $b=0.^\circ$22, in the second Galactic quadrant); 
and (c) Tombaugh~2 ($l=232.^\circ83$, $b=-6.^\circ$88, in the third Galactic quadrant).
They are old, distant clusters, and their properties are important to
understand the nature of the outer Galactic disc. In fact, they all lie in the region between 
about 12 and 16 kpc from the Galactic centre, where the radial metallicity distribution 
seems to change its slope (see Sect.~\ref{discussion}). 
Furthermore, To~2 has been connected with the Monoceros ring and the Canis Major 
overdensity \citep[e.g.,][see also Sect.~\ref{lit-to2}]{frinchaboy04,bellazzini04}.

Our paper is organized as follows: in Sect.~\ref{lit} we give a short description of what is already available on these 
clusters in literature, while in Sect.~\ref{data} we describe our data and the
reduction procedure. In Sect.~\ref{cmd} we present the CMDs and in Sect.~\ref{param} we
detail the cluster parameter derivation. Finally,  Sect.~\ref{discussion} is dedicated to a summary and discussion, in 
particular of the radial Galactocentric metallicity gradient and the importance of the three clusters in this context.

\begin{table*}
\begin{center}
\caption{ Stars in our Be~20 photometric catalogue for which RVs have been
published. FLAMES stands for the RVs from Sestito et al. (2007).  ID in first
column is our identificator; ID$_1$, ID$_2$ are in the  WEBDA system; ID$_3$ is
taken directly from the Frinchaboy et al. (2006) paper. 
The precision of the velocities is 0.5, 10, 1, 2 km~s$^{-1}$ for
RV$_0$,
RV$_1$, RV$_2$ and RV$_3$, respectively.
}
\begin{tabular}{rcccccccccccc}
\hline\hline
  ID  &  V    & I     &RA(2000)     &DEC(2000)     &FLAMES &\multicolumn{2}{c}{Friel et al. 2002}
  &\multicolumn{2}{c}{Yong et al. 2005} &\multicolumn{2}{c}{Frinchaboy et al. 2006}  & member \\
      &       &       & hh:mm:ss    &dd:pp:ss      & RV$_0$ &ID$_1$ &RV$_1$ &ID$_2$ &RV$_2$   &ID$_3$    &RV$_3$    &       \\
\hline      
  822 &14.792 &13.325 &05:32:37.957 &+00:11:09.61  &     &-- & --  &  5 & 77.3 & 10770 & 75.5 &   Y  \\  
  843 &15.145 &13.749 &05:32:38.960 &+00:11:20.33  &78.6 &-- & --  &  8 & 78.9 &  --   &   -- &   Y  \\ 
   28 &15.750 &14.565 &05:32:39.228 &+00:10:31.04  &     &10 & 82  & -- &  --  & 10810 & 63.7 &   Y  \\ 
  354 &16.922 &15.752 &05:32:36.886 &+00:11:49.49  &     &22 & 84  & 22 & 78.9 &  --   & 73.7 &   Y  \\ 
  185 &17.080 &15.928 &05:32:33.213 &+00:09:34.79  &     &27 & 15  & -- &  --  & 10617 &  0.1 &   N  \\ 
  395 &16.841 &16.023 &05:32:42.328 &+00:12:29.39  &     &29 & 61  & -- &  --  &  --   &   -- &   Y  \\ 
  368 &17.446 &16.312 &05:32:37.014 &+00:12:00.54  &     &33 & 54  & -- &  --  & 10741 & 74.5 &   Y  \\ 
  281 &17.543 &16.442 &05:32:33.276 &+00:10:53.32  &     &37 & 69  & -- &  --  &  --   &   -- &   Y  \\ 
  780 &17.496 &16.492 &05:32:33.509 &+00:11:37.62  &     &39 & 69  & -- &  --  &  --   &   -- &   Y  \\ 
  137 &17.613 &16.534 &05:32:46.573 &+00:08:41.92  &     &-- & --  & -- &  --  & 10940 & 40.2 &   N  \\ 
  396 &19.291 &17.687 &05:32:41.827 &+00:12:29.98  &     &-- & --  & -- &  --  & 10864 & 30.8 &   N  \\ 
   72 &17.040 &15.960 &05:32:34.739 &+00:15:38.72  &     &-- & --  & -- &  --  & 10645 & 24.4 &   N  \\ 
  179 &16.939 &16.205 &05:32:39.000 &+00:09:27.31  &     &-- & --  & -- &  --  & 10806 & 35.3 &   N  \\ 
   48 &16.584 &15.697 &05:32:50.012 &+00:11:54.12  &     &-- & --  & -- &  --  & 10979 & 32.5 &   N  \\ 
  991 &19.064 &17.739 &05:32:50.735 &+00:16:12.99  &     &-- & --  & -- &  --  & 10980 & 53.8 &   N  \\ 
   74 &15.919 &14.673 &05:32:46.420 &+00:15:52.23  &85.4 &-- & --  & -- &  --  &  --   &	-- &   N? \\
  702 &16.187 &14.943 &05:32:36.774 &+00:11:04.84  &78.5 &11 & --  & -- &  --  &  --   &	-- &   Y  \\
  977 &15.447 &13.992 &05:32:50.067 &+00:16:16.59  &37.7 &-- & --  & -- &  --  &  --   &	-- &   N  \\
\hline					        
\end{tabular}				        
\label{rvbe20}
\end{center}
\end{table*}

\begin{table*}
\begin{center}
\caption{ RVs for the two stars  (both members) in Be~66 analysed by Villanova et al. (2005). ID, magnitudes, and coordinates are from our work.}
\begin{tabular}{rcccccccccccc}
\hline\hline
  ID & ID$_{\rm WEBDA}$ & B     &  V    & I     & RA(2000)    & DEC(2000)   & RV             \\
      &      		&	&	&	& hh:mm:ss    & dd:pp:ss    & km s$^{-1}$    \\
 \hline     
 6521  & 785 	&20.375 &18.232 &15.596 & 03:04:02.90 & +58:43:57.0 &-50.7 $\pm$ 0.1 \\
 5793  & 934  	&20.357 &18.217 &15.595 & 03:04:06.41 & +58:43:31.0 &-50.6 $\pm$ 0.3 \\
\hline					        
\end{tabular}				        
\label{rvbe66}
\end{center}
\end{table*}

\begin{table*}
\centering
\caption{ RVs for stars in To~2. ID, magnitudes, and coordinates are from our
work. RV$_1$: Brown et al 1996; RV$_2$: Friel et al. 2002; RV$_3$: Frinchaboy et al. 2008; ID$_4$
and RV$_4$:  Villanova et al. (2010) (their Tables 1 and 5). } 
\begin{tabular}{r c rrr cc rrr rr c}
\hline\hline
 ID & ID$_{\rm WEBDA}$ & B  & V &  I	  & RA (2000)	& Dec (2000) &RV$_1$ &RV$_2$ &RV$_3$  &ID$_4$ & RV$_4$  &Member? \\
      &      &  & &        &  hh:mm:ss   & dd:pp:ss     &km~s$^{-1}$ &km~s$^{-1}$ &km~s$^{-1}$  & &km~s$^{-1}$ &\\
\hline
4736   & 0007  &13.974   &12.893   &11.654    &07:02:58.281  &-20:43:54.15  & 117.2 &		  &	 &	&	& Y \\
2272   & 0017  &15.293   &13.401   &11.280    &07:03:06.329  &-20:50:07.95  & 113.9 &  120.0  &      &      &	    & Y \\
2274   & 0020  &14.238   &13.628   &12.944    &07:03:20.398  &-20:47:11.41  &	    &  123.0  &      &      &	    & Y \\
4745   & 0022  &14.443   &13.850   &13.154    &07:03:26.967  &-20:46:00.78  & 117.2 &  123.0  &      &      &	    & Y \\
2276   & 0023  &15.588   &13.831   &11.886    &07:03:14.100  &-20:52:47.41  &	    &	63.0  &      &      &	    & N \\
4746   & 0024  &14.557   &13.967   &13.245    &07:03:05.909  &-20:44:18.48  &	    &  117.0  &      &      &	    & Y \\
2278   & 0028  &14.672   &14.050   &13.277    &07:03:05.081  &-20:52:35.09  &	    &  132.0  &      &      &	    & Y \\
2348   & 0047  &15.210   &14.662   &13.967    &07:03:24.512  &-20:45:34.34  &	    &  116.0  &      &      &	    & Y \\
2349   & 0048  &15.317   &14.672   &13.932    &07:02:48.134  &-20:49:54.55  &	    &  109.0  &      &      &	    & Y \\
2358   & 0056  &16.044   &14.865   &13.591    &07:03:00.935  &-20:45:21.94  &	    &  131.0  &      &      &	    & Y \\
2356   & 0057  &15.378   &14.845   &14.200    &07:02:47.275  &-20:52:59.40  &	    &  142.0  &      &      &	    & Y \\
2360   & 0058  &15.930   &14.897   &13.693    &07:03:00.169  &-20:45:18.48  &	    &  117.0  &      &      &	    & Y \\
2293   & 0061  &15.799   &15.058   &14.273    &07:02:59.391  &-20:50:52.42  &	    &  139.0  &      &      &	    & Y \\
2294   & 0062  &15.537   &15.073   &14.530    &07:03:10.992  &-20:48:29.87  &	    &  129.0  &      &      &	    & Y \\
2174   & 0063  &16.413   &15.062   &13.590    &07:03:06.851  &-20:48:28.12  &	    &	      & 121.0&      &	    & Y \\
2385   & 0076  &16.622   &15.316   &13.898    &07:03:23.050  &-20:53:38.89  &	    &	      &  84.9&      &	    & N \\
2388   & 0080  &16.728   &15.355   &13.816    &07:03:24.744  &-20:48:13.32  &	    &	      &  54.2&  472 &  53.6 & N \\
2299   & 0089  &16.680   &15.468   &14.036    &07:03:15.411  &-20:52:25.82  &	    &	      &  90.6&  169 &  90.0 & N \\
2402   & 0096  &16.840   &15.565   &14.113    &07:03:20.364  &-20:45:55.57  &	    &	      & 127.7&      &	    & N \\
2180   & 0097  &16.982   &15.552   &13.935    &07:03:19.759  &-20:51:52.41  &	    &	      & 105.8&      &	    & N \\
2158   & 0098  &16.919   &15.557   &14.014    &07:03:09.236  &-20:51:28.99  &	    &	      & 121.7&      &	    & Y \\
2165   & 0102  &16.646   &15.689   &14.366    &07:03:06.566  &-20:49:36.70  &	    &	      & 122.0&      &	    & Y \\
2412   & 0106  &16.744   &15.675   &14.494    &07:03:21.766  &-20:54:08.13  &	    &	      &  43.4&      &	    & N \\
2413   & 0107  &16.800   &15.677   &14.394    &07:02:53.880  &-20:46:39.11  &	    &	      &  93.2&      &	    & N \\
2417   & 0109  &16.992   &15.732   &14.372    &07:03:13.710  &-20:52:58.44  &	    &	      &  74.4&  160 &  74.5 & N \\
2183   & 0110  &17.049   &15.731   &14.242    &07:03:06.117  &-20:50:30.77  &	    &	      & 122.1&  299 & 121.8 & Y \\
1463   & 0115  &16.856   &15.769   &14.565    &07:03:12.533  &-20:49:26.82  &	    &	      &  39.0&      &	    & N \\
 270   & 0124  &17.184   &15.904   &14.460    &07:03:07.289  &-20:50:00.82  &	    &	      & 121.8&      &	    & Y \\
2434   & 0126  &17.065   &15.911   &14.579    &07:02:55.195  &-20:49:20.80  &	    &	      & 120.6& 1886 & 120.0 & Y \\
2436   & 0127  &17.188   &15.947   &14.526    &07:02:55.478  &-20:51:15.43  &	    &	      & 121.1&  238 & 120.5 & Y \\
2439   & 0135  &17.169   &15.961   &14.551    &07:02:53.928  &-20:50:09.69  &	    &	      & 119.2& 1672 & 119.1 & Y \\
 693   & 0140  &17.215   &15.974   &14.572    &07:02:59.738  &-20:49:33.36  &	    &	      & 121.2&      &	    & Y \\
2441   & 0141  &17.119   &15.969   &14.719    &07:02:51.115  &-20:47:15.09  &	    &	      & 104.0& 2343 & 104.0 & N \\
2446   & 0142  &17.132   &15.990   &14.734    &07:03:27.926  &-20:52:19.78  &	    &	      & 101.1&      &	    & N \\
2450   & 0146  &17.207   &16.051   &14.684    &07:03:09.695  &-20:45:48.97  &	    &	      & 119.5&      &	    & Y \\
2463   & 0158  &17.273   &16.147   &14.799    &07:03:07.768  &-20:46:17.21  &	    &	      & 118.8&      &	    & Y \\
 755   & 0162  &17.273   &16.123   &14.836    &07:03:01.582  &-20:47:59.53  &	    &	      & 118.0&      &	    & Y \\
 556   & 0164  &17.347   &16.170   &14.759    &07:03:20.483  &-20:46:47.57  &	    &	      & 122.6&  591 & 125.1 & Y \\
 824   & 0165  &17.364   &16.169   &14.776    &07:03:03.513  &-20:48:48.18  &	    &	      & 117.2&      &	    & Y \\
 268   & 0177  &17.379   &16.243   &14.876    &07:03:07.204  &-20:50:20.41  &	    &	      & 118.7&      &	    & Y \\
 424   & 0179  &17.411   &16.243   &14.897    &07:03:13.240  &-20:49:42.05  &	    &	      & 123.7&      &	    & Y \\
1601   & 0182  &17.404   &16.256   &14.928    &07:03:02.643  &-20:48:23.60  &	    &	      & 120.9& 3763 & 120.1 & Y \\
2474   & 0190  &17.331   &16.285   &15.099    &07:02:52.319  &-20:44:38.17  &	    &	      &  80.6&      &	    & N \\
 205   & 0191  &17.463   &16.309   &14.927    &07:03:05.082  &-20:48:57.53  &	    &	      & 121.2&      &	    & Y \\
2477   & 0192  &17.462   &16.301   &14.911    &07:02:53.302  &-20:48:01.26  &	    &	      &  96.3& 2185 &  96.3 & N \\
 246   & 0196  &17.557   &16.338   &14.892    &07:03:06.471  &-20:49:16.79  &	    &	      & 123.6&      &	    & Y \\
 164   & 0199  &17.516   &16.336   &14.955    &07:03:04.021  &-20:49:07.82  &	    &	      & 121.5& 3574 & 121.7 & Y \\
2489   & 0201  &17.476   &16.362   &15.046    &07:02:57.400  &-20:52:58.15  &	    &	      &  57.6&      &	    & N \\
1974   & 0207  &17.582   &16.402   &14.944    &07:03:19.393  &-20:48:39.53  &	    &	      &  32.0&      &	    & N \\
 364   & 0215  &17.641   &16.488   &15.085    &07:03:11.229  &-20:49:33.02  &	    &	      & 128.8&      &	    & N \\
 320   & 0217  &17.491   &16.559   &15.299    &07:03:09.119  &-20:49:25.75  &	    &	      & 151.9&      &	    & N \\
2499   & 0219  &17.570   &16.430   &15.137    &07:02:48.855  &-20:44:10.78  &	    &	      & 134.1& 2822 & 133.5 & N \\
2507   & 0223  &17.576   &16.497   &15.274    &07:02:48.387  &-20:47:37.47  &	    &	      &  70.6&      &	    & N \\
2522   & 0231  &17.802   &16.584   &15.147    &07:03:28.589  &-20:46:26.35  &	    &	      & 122.1&      &	    & Y \\
  47   &       &18.219   &17.140   &15.833    &07:02:58.677  &-20:50:56.07  &	    &	      &      & 1512 & 117.6 & Y \\
 693   &       &17.215   &15.974   &14.572    &07:02:59.738  &-20:49:33.36  &	    &	      &      & 1827 & 120.9 & Y \\
 343   &       &17.204   &16.062   &14.759    &07:03:10.339  &-20:47:59.24  &	    &	      &      & 2184 & 119.8 & Y \\
2465   &       &17.355   &16.188   &14.831    &07:03:27.495  &-20:51:15.08  &	    &	      &      &  236 & 121.0 & Y \\
2469   &       &17.419   &16.244   &14.997    &07:02:58.142  &-20:44:02.69  &	    &	      &      & 2846 & 122.9 & Y \\
1398   &       &17.849   &16.710   &15.363    &07:03:04.617  &-20:48:12.86  &	    &	      &      & 3836 & 120.9 & N \\
\hline
\end{tabular}
\label{rvto2}
\end{table*}

\setcounter{table}{2}
\begin{table*} 
\centering
\caption{ (continuation)}
\begin{tabular}{r r rrr cc rrr rr c}
\hline\hline
 ID & ID$_{\rm WEBDA}$ & B  & V &  I	  & RA (2000)	& Dec (2000) &RV$_1$ &RV$_2$ &RV$_3$  &ID$_4$ & RV$_4$  &Member? \\
      &      &  & &        &  hh:mm:ss   & dd:pp:ss     &km~s$^{-1}$ &km~s$^{-1}$ &km~s$^{-1}$  & &km~s$^{-1}$ &\\
\hline
 215	   &   &17.966   &16.992   &15.777    &07:03:05.211  &-20:52:45.36  &	    &	      &      &  153 &  61.1 & N \\
2394	   &   &16.754   &15.441   &13.946    &07:02:41.994  &-20:50:51.89  &	    &	      &      & 1530 &  98.9 & N \\
2421	   &   &16.990   &15.781   &14.444    &07:03:33.150  &-20:50:30.41  &	    &	      &      & 1573 &  59.8 & N \\
2289	   &   &16.234   &14.748   &13.105    &07:02:59.724  &-20:49:49.49  &	    &	      &      & 1755 & 126.9 & N \\
2404	   &   &17.059   &15.583   &14.018    &07:03:34.302  &-20:46:52.81  &	    &	      &      & 2403 &  64.8 & N \\
1435	   &   &18.156   &17.139   &16.045    &07:03:09.446  &-20:51:01.66  &	    &	      &      &  250 &  87.2 & N \\
2677	   &   &18.330   &17.373   &16.301    &07:03:26.870  &-20:46:16.91  &	    &	      &      & 2510 &  18.3 & N \\
2425	   &   &17.068   &15.819   &14.457    &07:03:32.502  &-20:50:48.01  &	    &	      &      &  266 & 130.9 & N \\
2609	   &   &18.011   &17.022   &15.888    &07:03:20.741  &-20:44:06.15  &	    &	      &      & 2827 &  38.5 & N \\
2396	   &   &16.715   &15.480   &14.153    &07:02:42.796  &-20:43:33.07  &	    &	      &      & 2911 & 109.1 & N \\
2422	   &   &16.940   &15.791   &14.527    &07:02:58.306  &-20:43:04.78  &	    &	      &      & 2975 & 105.6 & N \\
  73	   &   &18.148   &16.987   &15.667    &07:03:00.315  &-20:46:42.59  &	    &	      &      & 3332 & 130.6 & N \\
 185	   &   &17.107   &15.860   &14.448    &07:03:04.458  &-20:49:18.20  &	    &	      &      & 3562 &  51.3 & N \\
 745	   &   &17.175   &16.041   &14.696    &07:03:01.333  &-20:49:26.93  &	    &	      &      &  390 & 114.9 & N \\
2459	   &   &17.249   &16.132   &14.785    &07:03:23.382  &-20:46:49.24  &	    &	      &      &  589 & 135.0 & N \\
2418	   &   &16.917   &15.736   &14.396    &07:03:33.972  &-20:46:00.33  &	    &	      &      &  650 & 107.2 & N \\
2437	   &   &17.135   &15.922   &14.638    &07:02:42.171  &-20:46:05.23  &	    &	      &      &  651 & 121.7 & Y \\
\hline
\end{tabular}
\label{rvto2}
\end{table*}

\section{The three clusters in the literature} \label{lit}

The three clusters have already been partly studied in the past and we 
briefly summarise here the available information.
We have retrieved (from the WEBDA\footnote{{\em http://www.univie.ac.at/webda/webda.html}, 
see \cite{merm03}} or the original papers) the values of radial velocity (RV) for stars in our catalogues. 
They are presented in Tables~\ref{rvbe20}, \ref{rvbe66}, and \ref{rvto2}, 
together with photometric data, coordinates, and identifications.
The information on membership and metallicity (see next) for the clusters will be used
in the present paper to help in the selection of the best-fitting synthetic CMDs. 

\subsection{Be~20} \label{lit-be20}
The first
calibrated photometry for Be~20 was presented by \cite{macminn}, who obtained $V, I$  data
on a 5.1$\times$5.1 arcmin$^2$ field using the KPNO 2.1m telescope. Their CMD 
is well defined, but shows an apparent lack of red clump (RC) stars. 
\cite{macminn} derived, using isochrones, an age of 6
Gyr, [Fe/H]$\simeq-0.23$, $(m-M)_V\simeq15.0$ and $E(V-I)\simeq0.16$. They
deduced a radius of 1.5 arcmin and a mass of about 1000 \msun. They considered
Be~20 worth of further interest also because of its large Galactocentric distance
(R$_{GC}\simeq15.8$ kpc) and unusual position below the Galactic plane (about
2.5 kpc).

\cite*{d01} presented $B,V,R,I$ data obtained at the 104-cm Naini Tal State
Observatory over a 6$\times$6 arcmin$^2$ field. At variance with the other photometric
works, they do think they see a horizontal branch in the innermost region of
Be~20. Using two different sets of isochrones they concluded for an age of about
5 Gyr, $E(B-V)=0.10$, Z=0.008 (i.e. [Fe/H]$\simeq-0.3$), $(m-M)_V\simeq15.1$, a
cluster radius of about 2.5 arcmin and a  R$_{GC}\simeq17.1$ kpc.

\cite{momany} showed results on $B,V$ photometry obtained with the Wide Field
Imager (WFI@2.2m ESO-Max Planck telescope) on a much larger field of view (about
30$\times$30 arcmin$^2$). These observations were intended to produce catalogues
over fields of view appropriate to the FLAMES fiber spectrograph and the authors
did not really  discuss the cluster properties.

Radial velocities (see Table~\ref{rvbe20}) and abundances based on low resolution spectroscopy
were presented by \cite{friel02} for nine stars; six of them appear to be 
cluster members and have $\langle RV \rangle=+70 \pm 13$ km~s$^{-1}$, [Fe/H]=$-0.61
\pm 0.14$ dex.
\cite{frinchaboy06} observed 20 objects in Be~20 at intermediate resolution and
derived precise RVs; only five stars resulted members and their average RV is
$+75.7 \pm 2.4$ km~s$^{-1}$.

Abundances based on high resolution spectroscopy were first measured by
\cite*{yong05}. They observed four stars, two in common with \cite{friel02}, one with
\cite{frinchaboy06}; their average RV is $+78.9 \pm 0.7$ km~s$^{-1}$.
Abundance analysis was possible only for two stars,
located near the RGB tip, giving [Fe/H]$\simeq-0.49 \pm 0.06$ dex.
\cite{yong05} derived detailed abundances and discussed this cluster in the framework
of Galactic abundance trends and different origin/population.
Finally, \cite{sestito08} obtained FLAMES/UVES spectra of six
stars in the field of Be~20 over the large WFI area; two of them are
confirmed cluster members, and their analysis gives a metallicity
[Fe/H]=$-0.3$ (rms=0.02) dex. 

\subsection{Be~66} \label{lit-be66}
For Be~66, the only available photometry reaching the main sequence Turn-Off  
(MSTO) is by \cite{pj96}. They observed a 5.1$\times$5.1 arcmin$^2$ field using the KPNO
2.1m telescope with the $V,I$ filters, obtaining a well  defined CMD. They
derived the following parameters: age = $3.5\pm1.0$ Gyr, $-0.23 \le 
{\rm [Fe/H]}\le
0$, $E(V-I)=1.60 \pm 0.05$, $(m-M)_V=17.40 \pm 0.20$ (implying a Galactocentric
distance R$_{GC}\simeq12.9$ kpc), radius of 1.2-3.5 arcmin and minumum mass of
$\sim$750 \msun; they also suggested the possibility of differential reddening.

\cite{villanovabe66} obtained high resolution spectra  of two red clump stars with
the HIRES spectrograph  on the Keck I telescope. They have RVs  $-50.6$ and
$-50.7$ km~s$^{-1}$ (see Table~\ref{rvbe66}),  and seem to be cluster members.
Abundance analysis was possible only for one star; Villanova et al. estimated 
[Fe/H]$=-0.48\pm0.24$ and claimed that other elements have solar scaled ratios. 
 
\subsection{To~2} \label{lit-to2}

Although with significant discrepancies, all authors concur that To~2 is a distant, old, rather metal-poor open cluster.
To~2 was discovered by \cite{tombaugh} and its first CMD, barely reaching the main sequence
turn-off, was published by \cite{adler_janes}. $B,V,I$ images were collected with a variety of cameras at the 1-m and 2.5-m telescopes in Las Campanas by \cite{k92}, who intended to find variable stars.
Their CMDs are of very good quality (see also Sect.~\ref{data_to2}) and they derive a distance of $6.3\pm0.9$ kpc and an age of 4 Gyr, assuming $E(V-I)=0.4$ and a metallicity one tenth of solar. \cite{pjm94} included To~2 in their list of old OCs, assigning it a value of $\delta V=1.5$, i.e., an age of
about 2.5 Gyr \citep[following the formula in][]{jp94}. No further photometric catalogues are 
freely available, but \cite{villanovato2} re-determined the parameters of To~2 on the basis of spectroscopy (see below) and photometry, deriving  an age of 2 Gyr, $(m-M)_V=15.1$, $E(B-V) = 0.25$,
with [Fe/H]=$-0.32$ dex (see next). 

The metallicity of this cluster has been determined using low-resolution spectroscopy by \cite{friel02},
who find [Fe/H]$= -0.4\pm0.09$ using 12 stars; the corresponding RVs are given in Table~\ref{rvto2}, together with the ones by others, for all stars in common with our photometric catalogue.
\cite{brown} analysed high-resolution spectra of
three stars, finding an average [Fe/H]$=-0.4\pm0.25$ (with $E(B-V)=0.4$), and rather normal
elemental ratios (e.g., slightly enhanced $\alpha$-elements, slightly deficient oxygen) although with a caveat on the uncertainties.

This cluster could even be of extragalactic origin, and only recently accreted by the Milky Way:
on the basis of positional and kinematical arguments, \cite{frinchaboy04} proposed that To~2 is part, together with other OCs, of  the Monoceros stream, 
also called GASS, Great Anticenter Stellar Stream \citep{newberg02,ibata03}, possibly due to a dissolving, merging galaxy.
\cite{bellazzini04} associated To~2 directly to the dissolving dwarf galaxy proposed as origin of the Monoceros stream \citep{martin04}, i.e., to Canis Major (CMa),  whose reality has however been questioned by others \citep[e.g.,][]{momanycma}.  To~2 has then attracted the attention of several groups and two papers on its metallicity have recently appeared.

\cite{frinchaboy08} analysed high-resolution UVES and GIRAFFE VLT spectra of 40 stars, of which only about one half turned out to be members on the basis of the RVs. Their surprising results is the presence of two sub-populations, both apparently member of the cluster, one more metal-poor, $\alpha$-rich ([Fe/H$=-0.28$, [Ti/Fe]$=+0.36$, seven stars) and more centrally concentrated, the other more metal-rich, with solar-scaled $\alpha$ elements ([Fe/H$=-0.06$, [Ti/Fe]$=+0.02$, 11 stars) and more external. 
If confirmed, it would be the first case of chemical inhomogeneity in OCs and maybe of multiple stellar generations.
To~2 would in any case represent a very different case from globular clusters (GCs).  There, inhomogeneities, initially regarded as ''anomalies", have been found since a long time, but only in light elements, like C, N, O, Na, etc, whose abundance variations are (anti-)correlated \citep{araa04}. In contrast, the metallicity of GCs, described by [Fe/H], is homogeneous to better than 10 per cent \citep{carretta09}, with very few exceptions, like $\omega$ Centauri, or M22. While the presence of multiple stellar generations in (probably all) GCs is presently widely accepted and is based on both photometric and spectroscopic evidence \citep[see, e.g.,][]{bragaglia10}, the mass of GCs is much higher than that of OCs, and mass is certainly of paramount importance in shaping the destiny of a cluster.
To explain the peculiar case of To~2, \cite{frinchaboy08} proposed several alternative solutions: two  overlapping, or merged, clusters, multiple star formation periods, or the presence of both an open cluster  and a stream,  remnant of the same dwarf galaxy where also To~2 was born.

This result has however been questioned by part of the same team; \cite{villanovato2} presented results on  GIRAFFE VLT spectra for 37 RGB and red clump stars (only 15 actually members, and only 13 with data good enough for the analysis). They found [Fe/H]$=-0.31\pm0.02$ dex (rms=0.07 dex) and {\em no} sign of a bimodal distribution.  All the other measured elements are also uniform. They discuss the cause of this very discrepant result and the only reasonable explanation appears to be the better quality of the new data and the fact that their spectra are in a redder region than the \cite{frinchaboy08} ones, hence more easily analysed for relatively metal-rich and cool (i.e., with very crowded spectra) stars as the RGB ones in To~2.
 
Whether  CMa, and perhaps the Monoceros ring, are truly of extragalactic origin, or are instead due to disc warping and flaring, or spiral arms, or a combination of these phenomena, and whether To~2 is truly associated to these structures, or is ``simply" a normal, Galactic-disc open cluster, the study of this part of the Galactic disc is important. Further observations, both photometric and spectroscopic, are then welcome, in particular to settle the question of the possible chemical inhomogeneity,

\section{Our data} \label{data}
The three clusters were observed from two observatories and with three telescopes; 
a log of the observations is given in Table 4.
The final photometric catalogues will be made available at the WEBDA.
They will present the $(B)VI_C$ magnitudes with errors, pixel coordinates, and
equatorial coordinates (tied to the GSC2 Catalogue using CataPack, a software
written by P. Montegriffo at the Bologna Observatory).

\begin{table*}
\begin{center}
\caption{Log of observations for the clusters and the control fields}
\setlength{\tabcolsep}{1mm}
\begin{tabular}{lccllllll}
\hline\hline
Field       & $\alpha_{2000}$ &$\delta_{2000}$   & UT Date & exptime B & exptime V & exptime I & Telescope & Istrument \\
 & & & (dd/mm/yy) & ~~~~~~(s) & ~~~~~~(s) & ~~~~~~(s) \\
\hline 
Be 20   &$05^h32^m34^s$ &$+00^\circ 11\arcmin ~00\arcsec$  & 24/11/2000
 & --
 & 900,  120,  20 
 & 900,  120,  20 
 & TNG & DOLORES \\
Be 20 - control   &$05^h32^m34^s$ &$+00^\circ 30\arcmin ~01\arcsec$ &  24/11/2000  
 &  -- 
 & 600, 120, 20 
 & 480, 120, 20 
 & TNG & DOLORES \\
\hline
Be 66   &$03^h04^m06^s$ &$+58^\circ 44\arcmin ~40\arcsec$ & 03/10/2000 
 & 1300, 600, 300, 60  
 & 2$\times$900, 60, 10 
 & 900, 60, 10 
& TNG & DOLORES \\
Be 66 -  control  &$03^h04^m20^s$  &$+58^\circ 24\arcmin ~05\arcsec$ &  24/11/2000
 &1800, 240, 30 
 & 900, 120, 20 
 & 120, 20 
& TNG & DOLORES \\ 
\hline
To 2   &$07^h02^m57^s$ &$-20^\circ 49\arcmin ~53\arcsec$ & 07-08/03/1995 
 & 1500, 300 
 & 2$\times$720, 300, 2$\times$60 
 & 2$\times$720, 243, 60 
 & Danish & Direct camera\\
  & & & 15/05/2001 
 & 200, 2$\times$10 
 & 150, 10 
 & 120, 10 
& Danish & DFOSC\\
  & & & 15/01/2002 
 & 1800 
 & 1200 
 & 2$\times$900 
 & NTT & SuSI2\\
To 2 -  control  &$07^h06^m06^s$  &$-20^\circ 50\arcmin ~20\arcsec$ &  15-16/05/2001
 & --
 & 600, 60, 10 
 & 500, 60, 10 
 & Danish & DFOSC\\ 
\hline
\end{tabular}
\end{center}
\label{logoss}
\end{table*}

\subsection{Be~20 and Be~66} \label{data20_66}

Observations of the two clusters and the two  control fields were  obtained
at the Telescopio Nazionale Galileo (TNG) on Canary Islands, during two nights 
in October and November 2000. We used DOLORES (Device Optimized for LOw RESolution), 
characterized by a field of view of 9.4$\times$9.4 arcmin$^2$ and a scale 
of 0.275\arcsec /pixel.
Since in these two runs we also observed Be~17 \citep{bra06a} and NGC~6939
\citep{andreuzzi}, we refer to those publications for details on observation
and reduction procedures. Very briefly, we used the
package {\sc DAOPHOT-II}
\citep{stetson87,davis94} to obtain PSF magnitudes, we corrected them to
the same scale of aperture photometry and calibrated to the standard 
Johnson-Cousins system using Landolt's areas \citep{landolt92}. 
The calibration equations adopted for Be~20 and its  control field 
observed on the same night, are:

\[       V = v -0.0947 \times (v-i) +1.1917  ~~(rms=0.014) \]
\[       I_C = i -0.0060 \times (v-i) +0.7638  ~~(rms=0.025) \]

The calibration equations adopted for Be~66 are:

\[       B = b +0.0475 \times (b-v) +1.4211  ~~(rms=0.012) \]
\[       V = v -0.0959 \times (v-i) +1.2003  ~~(rms=0.012) \]
\[       I_C = i +0.0422 \times (v-i) +0.7248  ~~(rms=0.012) \]

The calibration equations adopted for the  control field
of Be~66 are:

\[       B = b +0.0421 \times (b-v) +1.4043  ~~(rms=0.012) \]
\[       V = v -0.0947 \times (v-i) +1.1917  ~~(rms=0.014) \]
\[       I_C = i + 0.0017 \times (v-i) + 0.7589 ~~(rms=0.024), \]

 where $b, v$, and $i$, are the aperture corrected instrumental magnitudes 
after correction also for extinction and exposure time, and $B, V$, and
$I_C$ are the output magnitudes, calibrated to the Johnson-Cousins standard
system.

Finally, the completeness level of our photometry was derived with  extensive
artificial stars experiments, as in previous papers of this project. The
completeness factors are shown in Table~\ref{compl20-66-to2}. From these
experiments we also derived the errors typically associated to each magnitude
level, that are used in Sect.~\ref{param} to build the synthetic CMDs.

We compared our photometry for Be~20 with the ones of the two papers
presenting the derivation of cluster parameters and readily available through the  WEBDA, 
i.e., to the \cite{macminn} and the \cite{d01} data.  

\begin{figure}
\centering
\includegraphics[bb=20 100 574 574, clip, scale=0.45]{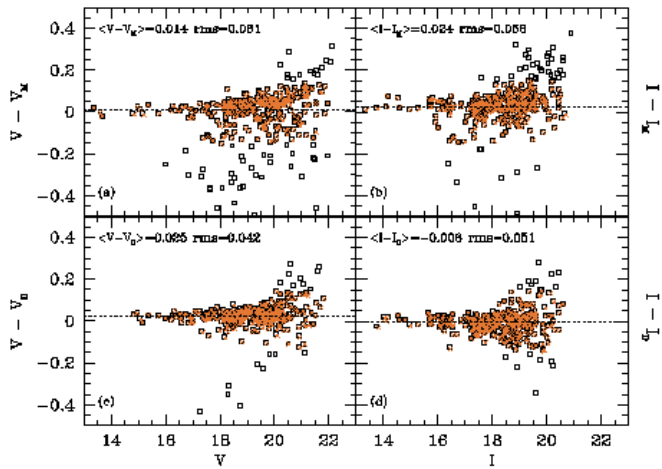} 
\caption{Comparison between our photometry and literature data for Be~20.
 Panels (a) and (b) show the comparison to the MacMinn et al. (1994) data; 
panels (c) and (d) show the comparison to the Durgapal et al. (2001) 
data.
}
\label{fig-conf1}
\end{figure}

We counter-identified stars in the different
catalogues and the results are presented in Fig.~\ref{fig-conf1}. 
In particular panels (a) and (b) in the figure show the comparison to the MacMinn et al. (1994) 
data, while panels (c) and (d) show the comparison to the Durgapal et al. (2001) data.
For each star in common between the three catalogs, V and I are the magnitudes 
in our photometry calibrated with the equations given in the text;
V$_M$ and I$_M$ are the corresponding magnitudes in the catalog of \cite{macminn} 
while V$_D$ and I$_D$ are the magnitudes of the corresponding stars in the catalog 
of \cite{d01}.

\begin{figure}
\centering
\includegraphics[bb=20 80 700 520, clip, scale=0.44]{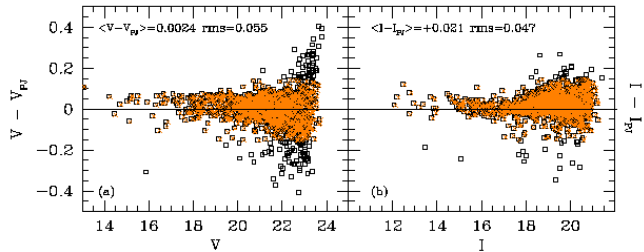} 
\caption{Comparison between our photometry and the one by Phelps \& Janes (1996)
 for Be~66.}
\label{fig-conf2}
\end{figure}

For Be~66, the comparison was done to \cite{pj96}, again only in
the $V$ and $I$ filters, since they did not obtain $B$ photometry.
The results are shown in Fig.~\ref{fig-conf2}. 
For each star in common between the two catalogs, V and I are the magnitudes 
in our photometry calibrated with the equations given in the text. 
V$_{PJ}$ and I$_{PJ}$ are the corresponding magnitudes in the catalog of \cite{pj96}.

For both clusters, the differences between the different photometries have been computed 
retaining only well counter-identified stars (corresponding to the subsamples indicated in colour 
in the electronic version of the figures), with a difference in magnitude less than 0.15 mag, in absolute value.
Differences are of the order of 0.02-0.03 mag in all cases.

\subsection{To~2} \label{data_to2}
Observations were obtained in La Silla, Chile, using two telescopes and three
different instruments: a direct CCD camera (6.4$\times$6.4 arcmin$^2$) in 1995
and DFOSC (Danish Faint Object Spectrograph and Camera, 13$\times$13 arcmin$^2$)
in 2001, both mounted at the  Danish telescope, and SuSI2 (Super Seeing Imager
2, 5$\times$5 arcmin$^2$), mounted at the New Tecnology Telescope (NTT) in 2002.
We observed a field centred on the cluster and one about 50 arcmin away 
as a comparison to separate cluster and field stars. 

The data reduction followed a standard procedure, see for instance \cite{difabrizio05}. 
Unfortunately, none of the nights turned out to be truly photometric, as we
found out by comparison to the published photometries by  \cite {k92} 
and \cite{pjm94}. We decided to calibrate our photometry to the one by
Kubiak et al., since it was the deepest one. No similar solution has been
possible for the comparison field; however, the photometric accuracy obtained
for it is enough for our goals (we only wish to separate cluster and field
stars), also because we only have quite short exposures for this field.

The results for the completeness tests are shown in Table~\ref{compl20-66-to2}.
The final catalogue for To~2 contains 6073 stars.
The limiting magnitude
depends on the distance from the centre, since we obtained only short
exposures with the widest field. In the following we will consider only
the central 5$\times$5 arcmin$^2$ region, observed with SuSI2 on NTT, to derive the cluster properties.

\begin{figure}
\centering
\includegraphics[bb=30 180 580 520, clip,scale=0.45]{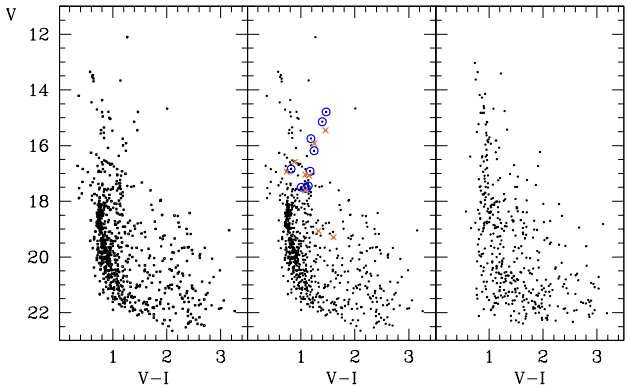} 
\caption{CMDs for Be~20 (left panel) and the control field (right panel).
In the middle panel we indicate with different symbols member (blue open
circles) and non member (orange crosses) stars, according to their RVs.
The latter are listed in Table~\ref{rvbe20}, together with the references.}
\label{fig-cmdbe20}
\end{figure}

\begin{figure}
\centering
\includegraphics[bb=40 30 577 577, clip, scale=0.45]{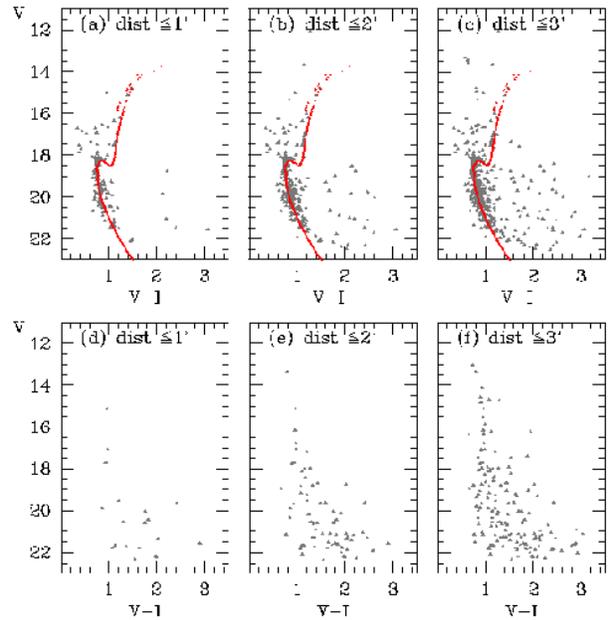} 
\caption{Radial CMDs for Be~20 (upper panels) and the control field (lower
panels). The stars within the 1\arcmin, 2\arcmin, 3\arcmin ~radii
are 136, 269, 401 for the cluster and 19, 70, 153 for the control
field, respectively. In the upper panels we show one of the best-fitting
solutions (see Sect.~\ref{param_be20}) to guide the eye.}
\label{fig-radbe20}
\end{figure}

\begin{table*}
\begin{center}
\caption{Completeness level for the clusters and control fields of Be~20 and Be~66 and for
the central field of To~2 as measured from the SuSI2 observations; mag is
the calibrated magnitude ($B$, $V$, or  $I_C$). 
}
\begin{tabular}{lllllllllllllllllll}
\hline\hline
mag & & c$_V$  &c$_I$  &&  c$_V$	&c$_I$
&& c$_B$  & c$_V$  &c$_I$   
&& c$_B$ & c$_V$ & c$_I$   
&& c$_B$  & c$_V$  &c$_I$       \\
      & & \multicolumn{2}{c}{Be 20} && \multicolumn{2}{c}{Be 20-control} 
       &&\multicolumn{3}{c}{Be 66} & & \multicolumn{3}{c}{Be 66-control}
       && \multicolumn{3}{c}{To 2} \\
\hline
 15.5 & & 1.00 & 1.00 && 1.00 & 1.00 && 1.00 & 1.00 & 1.00 && 1.00 & 1.00 & 1.00 && 1.00 & 1.00 & 1.00\\
 16.0 & & 1.00 & 0.96 && 0.98 & 0.99 && 1.00 & 1.00 & 1.00 && 1.00 & 1.00 & 0.99 && 1.00 & 1.00 & 1.00\\
 16.5 & & 0.98 & 0.93 && 0.98 & 0.96 && 1.00 & 1.00 & 1.00 && 1.00 & 1.00 & 0.99 && 1.00 & 1.00 & 1.00\\
 17.0 & & 0.94 & 0.91 && 0.97 & 0.96 && 1.00 & 1.00 & 1.00 && 1.00 & 1.00 & 0.97 && 1.00 & 1.00 & 1.00\\
 17.5 & & 0.94 & 0.88 && 0.95 & 0.94 && 1.00 & 1.00 & 1.00 && 1.00 & 0.99 & 0.97 && 1.00 & 1.00 & 1.00\\
 18.0 & & 0.92 & 0.85 && 0.91 & 0.98 && 1.00 & 1.00 & 1.00 && 1.00 & 0.98 & 0.95 && 1.00 & 1.00 & 1.00\\
 18.5 & & 0.91 & 0.77 && 0.91 & 0.93 && 1.00 & 0.96 & 0.94 && 1.00 & 0.98 & 0.94 && 1.00 & 1.00 & 1.00\\
 19.0 & & 0.91 & 0.67 && 0.90 & 0.88 && 1.00 & 0.97 & 0.92 && 0.99 & 0.95 & 0.93 && 1.00 & 1.00 & 1.00\\
 19.5 & & 0.88 & 0.47 && 0.86 & 0.76 && 0.99 & 0.96 & 0.89 && 0.99 & 0.98 & 0.85 && 1.00 & 1.00 & 1.00\\
 20.0 & & 0.79 & 0.27 && 0.75 & 0.53 && 0.97 & 0.96 & 0.83 && 0.96 & 0.92 & 0.60 && 1.00 & 1.00 & 1.00\\
 20.5 & & 0.68 & 0.08 && 0.55 & 0.30 && 0.96 & 0.95 & 0.78 && 0.94 & 0.86 & 0.32 && 1.00 & 1.00 & 0.97\\
 21.0 & & 0.46 & 0.00 && 0.31 & 0.10 && 0.97 & 0.94 & 0.67 && 0.92 & 0.85 & 0.13 && 1.00 & 1.00 & 0.82\\
 21.5 & & 0.23 & 0.00 && 0.13 & 0.00 && 0.95 & 0.93 & 0.43 && 0.92 & 0.71 & 0.03 && 1.00 & 1.00 & 0.80\\
 22.0 & & 0.10 & 0.00 && 0.03 & 0.00 && 0.92 & 0.86 & 0.19 && 0.90 & 0.48 & 0.00 && 0.93 & 1.00 & 0.75\\
 22.5 & & 0.02 & 0.00 && 0.01 & 0.00 && 0.92 & 0.69 & 0.00 && 0.80 & 0.24 & 0.00 && 0.85 & 0.95 & 0.68\\
 23.0 & & 0.01 & 0.00 && 0.00 & 0.00 && 0.80 & 0.38 & 0.00 && 0.56 & 0.11 & 0.00 && 0.81 & 0.85 & 0.38\\
 23.5 & & 0.00 & 0.00 && 0.00 & 0.00 && 0.50 & 0.03 & 0.00 && 0.30 & 0.02 & 0.00 && 0.79 & 0.77 & 0.14\\
 24.0 & & 0.00 & 0.00 && 0.00 & 0.00 && 0.16 & 0.00 & 0.00 && 0.15 & 0.00 & 0.00 && 0.76 & 0.77 & 0.10\\
 24.5 & & 0.00 & 0.00 && 0.00 & 0.00 && 0.00 & 0.00 & 0.00 && 0.04 & 0.00 & 0.00 && 0.70 & 0.68 & 0.03\\
 25.0 & & 0.00 & 0.00 && 0.00 & 0.00 && 0.00 & 0.00 & 0.00 && 0.00 & 0.00 & 0.00 && 0.65 & 0.50 & 0.00\\
 25.5 & & 0.00 & 0.00 && 0.00 & 0.00 && 0.00 & 0.00 & 0.00 && 0.00 & 0.00 & 0.00 && 0.39 & 0.39 & 0.00\\
 26.0 & & 0.00 & 0.00 && 0.00 & 0.00 && 0.00 & 0.00 & 0.00 && 0.00 & 0.00 & 0.00 && 0.16 & 0.20 & 0.00\\
 26.5 & & 0.00 & 0.00 && 0.00 & 0.00 && 0.00 & 0.00 & 0.00 && 0.00 & 0.00 & 0.00 && 0.03 & 0.03 & 0.00\\
 27.0 & & 0.00 & 0.00 && 0.00 & 0.00 && 0.00 & 0.00 & 0.00 && 0.00 & 0.00 & 0.00 && 0.00 & 0.00 & 0.00\\
\hline
\end{tabular}
\label{compl20-66-to2}
\end{center}
\end{table*}

\section{The colour - magnitude diagrams} \label{cmd}

\begin{figure}
\centering
\includegraphics[bb=40 0 580 520, clip,scale=0.45]{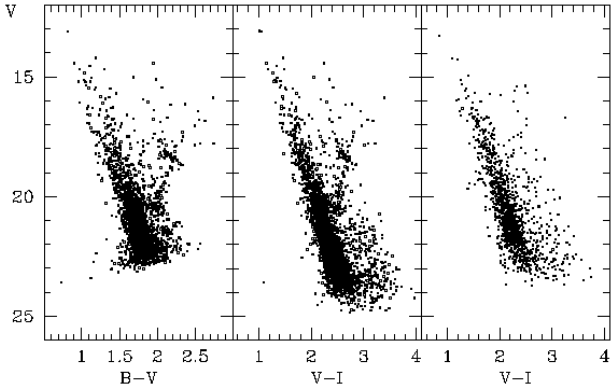} 
\caption{CMDs for Be~66 (left and middle panels) and the control field 
(right panel, shown only in the $V,V-I$ plane).
}
\label{fig-cmdbe66}
\end{figure}

\begin{figure}
\centering
\includegraphics[bb=50 0 570 700, clip, scale=0.45]{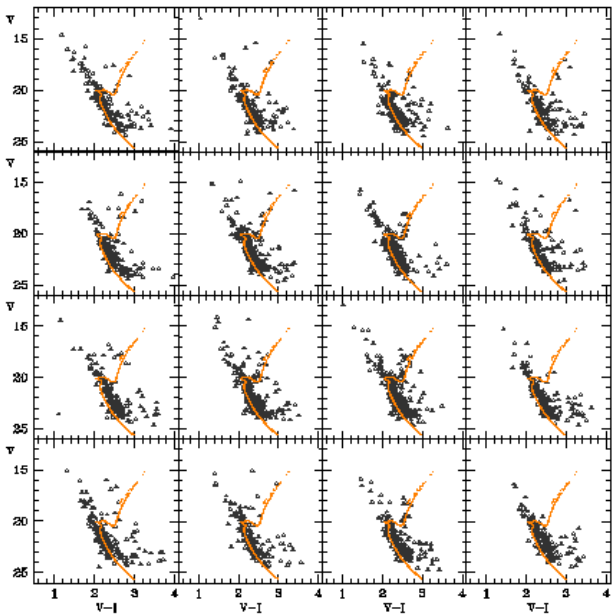}
\caption{CMDs obtained in different regions (500$\times$500 pixels wide,
i.e., 2.2$\times$2.2 arcmin$^2$) of Be~66. East is to the left and North
to the top, and the cluster is located at the
centre of the field of view.  The isochrone (FST models with age 3 Gyr and
metallicity Z=0.006) shown here represents one of the
best solutions (see Sect.~\ref{param}) and clearly indicates that differential
reddening is present, but only at a few per cent level. 
}
\label{fig-box}
\end{figure}

\begin{figure}
\centering
\includegraphics[bb=70 0 520 680, clip, scale=0.55]{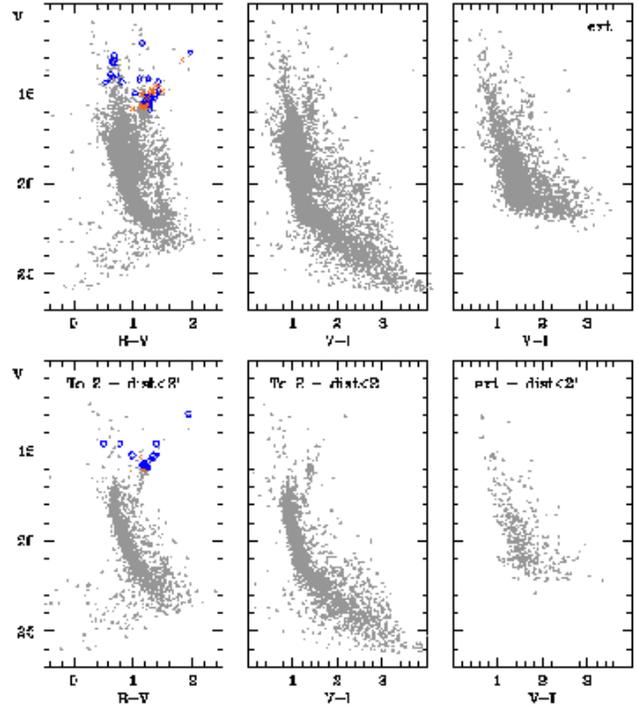}
\caption{$V, B-V$ and $V, V-I$ CMDs for To~2 and the comparison field. The
upper panels show the entire field, while the lower panels show only stars
within a 2 arcmin distance. We also indicate, only for the $V, B-V$ diagram,  
member (blue open circles) and non member (orange crosses)
stars, according to their RVs (see Table~\ref{rvto2}).}
\label{fig-cmdto2}
\end{figure}

\begin{figure}
\centering
\includegraphics[bb=80 420 470 680,clip, scale =0.55]{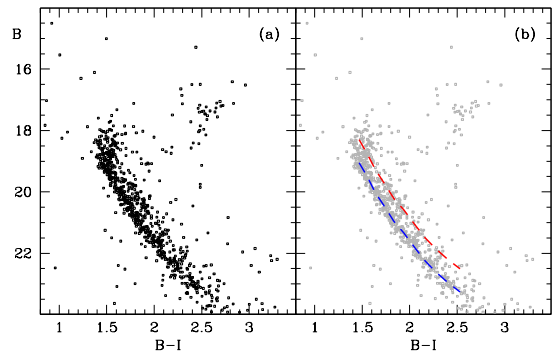}
\caption{(a) CMD in $B,B-I$ of the central region of To~2 
(within a radius of 1 arcmin)
showing the binary sequence. (b) The same, but showing the MS ridge line and
the same line shifted by 0.7 mag brighter.}
\label{fig-binto2}
\end{figure}

\begin{itemize}
\item  Be~20 --
Fig.~\ref{fig-cmdbe20} shows the CMD obtained for the field centred on Be~20 
(left panel) and for the  control field (right panel); the middle panel
indicates (with larger symbols) the stars for which information on membership is
available thanks to the RVs.
We clearly see the main sequence for about 4 magnitudes below the
MSTO, located around $V=18.2$, $V-I=0.7$. We also identify the
red clump with the few stars at  $V \simeq 16.0$, $V-I \simeq 1.2$.
Given the field stars distribution, stars above the MSTO may be attributed
to a blue stragglers population or to fore/background contamination.
Indeed, of the three stars with measured RV present in this region, two are
field objects and one is a cluster member. More could be said only after
decontamination, either statistical or (better) through actual measurement
of the individual membership status via RV or proper motion.

Fig.~\ref{fig-radbe20} shows the radial distribution of stars in the  cluster
field,  compared to an equal area in the  control field; this is very
useful to better understand which are the true cluster sequences and the
degree of field contamination (see also next Sect.).

\item  Be~66 --
The CMDs for Be~66 and the control field are presented in
Fig.~\ref{fig-cmdbe66}. Notice the very red colours of the sequences, due to the
high reddening, and the width of the evolutionary sequences.  \cite{jp94}
suspected the presence of differential reddening, a very reasonable possibility
given its high value. We estimate that a variation of only a few per cent 
(see Sect.\ref{param_be66}) is sufficient to justify the observed spread.
 Fig.~\ref{fig-box} shows a graphical representation of
this: the CMDs obtained in sub-regions of the frame are all similar to each
other and can be fit by the same isochrone (see Sect.~\ref{param_be66} 
for the choice of the
best one) once some difference in  reddening is accepted. We notice in
particular that the reddest CMDs correspond to the most eastern subregions in
the two central rows of Fig.~\ref{fig-box}; while the least reddened CMDs are
sparsely located: one in the bottom-left (south-east) panel and the other in the
third top panel from left (north-east). This distribution suggests a
clumpy/inhomogeneous extinction, rather than a systematic reddening variation.

In spite of the high dispersion and contamination, the main evolutionary phases,
MS, MSTO, SGB, RGB, clump and possibly AGB are quite well defined, with the MSTO at
$V \simeq 20.1$, $B-V \simeq 1.6$, and $V-I \simeq 2.1$, and the clump - RGB
intersection at $V \simeq 18.4$, $B-V \simeq 2.2$, and $V-I \simeq 2.7$.

\item  To~2 --
The situation for To~2 is more complicated, since the cluster was observed with
three different instruments with very different field of view, and to very
different depths (see Table 4). Fig.~\ref{fig-cmdto2} shows the
combination of all data for the central field in the upper row, left
and middle panels, for the $V,B-V$ and $V,V-I$ CMDs, respectively.
The entire control field is shown in the upper, right panel, only in $V,V-I$; we can
immediately appreciate the different depths reached by the various instruments.

To better separate the cluster CMD from the background, we plot in the lower
panel of Fig.~\ref{fig-cmdto2} only stars within a 2 arcmin distance from the
cluster centre (left and middle panels) and within the same area in the control
field (right panel). The evolutionary phases are
much better recognizable, with the MSTO at $V \simeq 17.5$, $B-V \simeq 0.60$,
and $V-I \simeq 0.65$, and the red clump
at $V \simeq 16.2$, $B-V \simeq 1.12$, and $V-I \simeq 1.33$. 

Member and non member stars,
according to their RVs (see Table~\ref{rvto2}), are indicated with different
symbols in the $V,B-V$ CMD. This information is very useful to confirm the
position of the RGB and red clump; unfortunately, RVs are not available for
stars on the MS and the MSTO. However, as found by 
\citet[][see their Sects. 3 and 6]{frinchaboy08}, 
even if the difference in RV between the cluster and the Galactic field stars is 
large enough to ensure a good decontamination, the same is not
valid for the GASS/Monoceros component
and some confusion may still be present.

If we further restrict to the stars within 1 arcmin from the cluster centre,
where the cluster stars dominate over the background, it is possible to see a
clear indication of the presence of binaries, as shown in
Fig.~\ref{fig-binto2}(a).
The secondary MS is visible above and to the red of the single-stars MS; we use
here $B-I$ to have a larger baseline in colour. In Fig.~\ref{fig-binto2}(b)
we indicate the two MSs; it is evident how the secondary, binary sequence
complicates the definition of the position of the single-stars MSTO.

\end{itemize}

\begin{figure}
\centering
\includegraphics[bb=50 180 550 510,clip, scale=0.45]{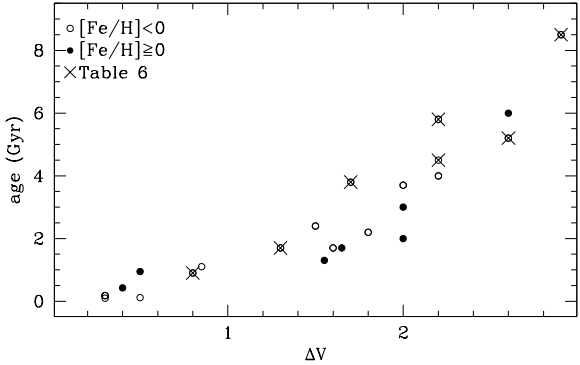}
\caption{$\delta V$ versus age for the 23 OCs presently available in the BOCCE
sample, 16 clusters from Bragaglia \& Tosi (2006), seven from Table 6. The 
meaning of the different symbols is indicated in the figure.}
\label{figdeltav}
\end{figure}

The difference in magnitute between 
the MSTO and the red clump ($\delta V$) can be used as an age indicator,
after a suitable calibration \citep[e.g.,][]{pj96,friel95,bt06}. This
may be useful especially when dealing with large samples and/or a photometry that does not reach much below the MSTO. 
In Table~\ref{deltav} we indicate these values (measured as described in \citealt{bt06}) for both the 
three OCs analysed here and four others we published after \cite{bt06}. We also give other relevant
information, like age and metallicity (obtained with the synthetic CMD technique and the BBC tracks, see Sect.~\ref{param})
and [Fe/H]. The last column gives references for the parameters derived from photometry (Cols. 2-4, all  by our group) 
and from high-resolution spectroscopy (Col. 5), respectively.
The references are: 1: \cite{bra06a}; 
2: \cite{friel05} ; 3: \cite{tosi07}; 4: \cite{sestito08};
5: this paper; 6: \cite{villanovabe66};
7: \cite{villanovato2}; 8: \cite{bra06b}. In Fig.~\ref{figdeltav} we plot the values of $\delta V$ versus age derived using stellar models for all OCs in the BOCCE sample.  The relation is clearly not a simple one, as witnessed by the spread, only 
a part of which is  due to errors. Metallicity plays surely a role,
see e.g., \cite{tat89} who decided to use a different indicator, combining differences in magnitude and colour, to take it into
account.  We do not derive here a relation between $\delta V$ and age, deferring the task to when more OCs will be available on
our scale.

\begin{table}
\centering
\caption{Values for $\delta V$, age in Gyr, Z, and [Fe/H] from high-resolution spectroscopy for the
seven clusters analysed after Bragaglia \& Tosi (2006). See text for the references.
}
\begin{tabular}{lccrcc}
\hline\hline 
Cluster   &$\delta V$ & age & Z~~~ &[Fe/H] & References \\
\hline
Be~17     & 2.9 & 8.5 & 0.008 & $-0.10$ & 1, 2 \\      
Be~32     & 2.6 & 5.2 & 0.008 & $-0.29$ & 3, 4 \\ 
Be~20     & 2.2 & 5.8 & 0.008 & $-0.30$ & 5, 4 \\ 
King~11   & 2.2 & 4.5 &  0.02 & --      & 3, - \\ 
Be~66     & 1.7 & 3.8 & 0.008 & $-0.48$ & 5, 6 \\ 
To~2      & 1.3 & 1.7 & 0.008 & $-0.31$ & 5, 7 \\ 
NGC~3960  & 0.8 & 0.9 &  0.02 & $-0.12$ & 8, 8 \\
\hline
\end{tabular}\label{deltav}
\end{table}

\section{Cluster parameters} \label{param}

As for all the clusters of the BOCCE project (see \citealt{bt06} and
references    therein), we have derived age, distance and reddening following
the synthetic  CMD method originally described by \cite{tosi91}. The best values
of the  parameters   are found by selecting the cases providing synthetic CMDs
with morphology,  colours, number of stars in the various evolutionary phases
and luminosity functions (LFs) in better agreement with the observational ones. 
To evaluate the goodness of the model predictions we quantitatively compare them 
with the observational LFs, stellar magnitude  and colour distributions, and number of 
objects at the MSTO, the clump and the RGB. 
As discussed by \cite{Kalirai04} and \cite{tosi07}, where the luminosity and colour distributions
of each model were independently compared with the data using a Kolmogorov-Smirnov test 
and a $\chi^2$ test, even sophisticated statistical procedures do not provide safer estimates 
of the cluster parameters. This is due to the background/foreground contamination and to 
the small number of objects usually measured in key evolutionary phases, such as the MSTO, 
the red clump and the SGB, fundamental in the identification of the cluster age. 
In the most favorable cases, the statistical tests confirm our choices of best synthetic models. 
In the cases of the three systems presented here, the parameter selection is even more uncertain, 
due to further decontamination problems related to the complicated and
inhomogeneous distribution of their fore/background interlopers. Yet, in these
unfavourable conditions the synthetic CMD approach is even safer (or less
unsafe) than isochrone fitting than in standard cases, thanks to its capability
to exploit all the available information on shape, population and position in
the CMD of the various evolutionary phases.
The method cannot provide strictly unique results, but allows to significantly reduce 
the range of acceptable parameters.

In our procedure, the synthetic stars, extracted from
the adopted stellar evolutionary tracks with a Monte Carlo approach taking into
accout the adopted initial mass function (IMF) and star formation 
law, are attributed the photometric error derived from 
the artificial stars tests performed on the actual images. For all the BOCCE
clusters we assume a Salpeter's IMF and a constant star formation lasting 5 Myr
from the epoch of activation. The extracted stars are
retained in (or excluded from) the synthetic CMD according to the photometry 
completeness  factors listed in Table~\ref{compl20-66-to2}.
We have computed the synthetic CMDs both with and without taking into account
the possible contribution from unresolved binaries (see \citealt{bt06} 
for a description of how binaries are included in the synthetic CMDs). 
Binary members are assumed to follow the same initial mass function 
(Salpeter's) as isolated stars and to have  
random mass ratio between primary and secondary components.
In no case were the CMDs 
without binaries in agreement with the obervational ones. All the CMDs 
discussed here assume that  30\%  of the 
cluster measured stars are actually unresolved binaries. 
This fraction is consistent with what we find for the majority of the BOCCE
clusters . 

As usual,  to test the effects of the adopted stellar evolution models on the
derived   parameters, we run the simulations with three different types of
stellar    tracks, with different assumptions for the treatment of convection,
opacities  and equation of state.  The adopted models are listed in
Table 7,  where the  corresponding references are also given, as well
as the  model metallicity and the information on their corresponding
overshooting assumptions.  

To estimate the metallicity which better reproduces the photometric properties 
of the cluster, we have created the  synthetic CMDs adopting, for each type of
stellar models, metallicities  ranging from solar down to Z=0.004. We still
assume as solar metallicity  models those with Z=0.02, both for consistency with
the BOCCE previous studies (see \citealt{bt06}) and because they are the ones 
calibrated on the Sun by their authors. We consider only as indicative the 
metallicities obtained with  our photometric studies and refer to high
resolution  spectroscopy for a safer determination of the chemical abundances.

The transformations from the theoretical luminosity
and effective temperature  to the Johnson-Cousins magnitudes and colours have
been performed using  the Bessell,  Castelli \& Pletz (1998 and private
communication) conversion tables of the metallicity of the adopted models. We
assume $A_V = (3.25+0.25\times(B-V)+0.05\times\ebv)\times\ebv$ 
and $E(V-I)=1.25 \times E(B-V) \times[1+0.06(B-V)_0 +0.014 E(B-V)]$ 
from Dean et al. (1978) for all sets of models. 
Using the same source for the conversion tables for all models, we can be 
confident that the
differences in the synthetic CMDs (and therefore in the cluster parameters)
resulting from different stellar models must be fully ascribed
to the intrinsic differences (input physics, opacities, etc.) of the
models themselves and not to different photometric conversions. 

\begin{table}
\begin{center}
\caption{Stellar evolution models adopted for the synthetic CMDs. The FST
models actually adopted here are an updated version of the published ones
(Ventura, private communication).}
\begin{tabular}{cccl}
\hline\hline
   Set  &metallicity & overshooting & Reference \\
\hline
BBC & 0.008 & yes 		    &Fagotto et al. 1994 \\
BBC & 0.004 & yes 		    &Fagotto et al. 1994 \\
BBC & 0.02  & yes 		    &Bressan et al. 1993 \\
FRA & 0.006 & no  		    &Dominguez et al. 1999 \\
FRA & 0.01  & no  		    &Dominguez et al. 1999 \\
FRA & 0.02  & no  		    &Dominguez et al. 1999 \\
FST & 0.006 & $\eta$=0.00,0.02,0,03 &Ventura et al. 1998\\
FST & 0.01  & $\eta$=0.00,0.02,0,03 &Ventura et al. 1998\\
FST & 0.02  & $\eta$=0.00,0.02,0,03 &Ventura et al. 1998\\
\hline
\end{tabular}
\end{center}
\label{models}
\end{table}

\subsection{Be~20} \label{param_be20}

\begin{figure*}
\centering
\includegraphics[scale=0.65]{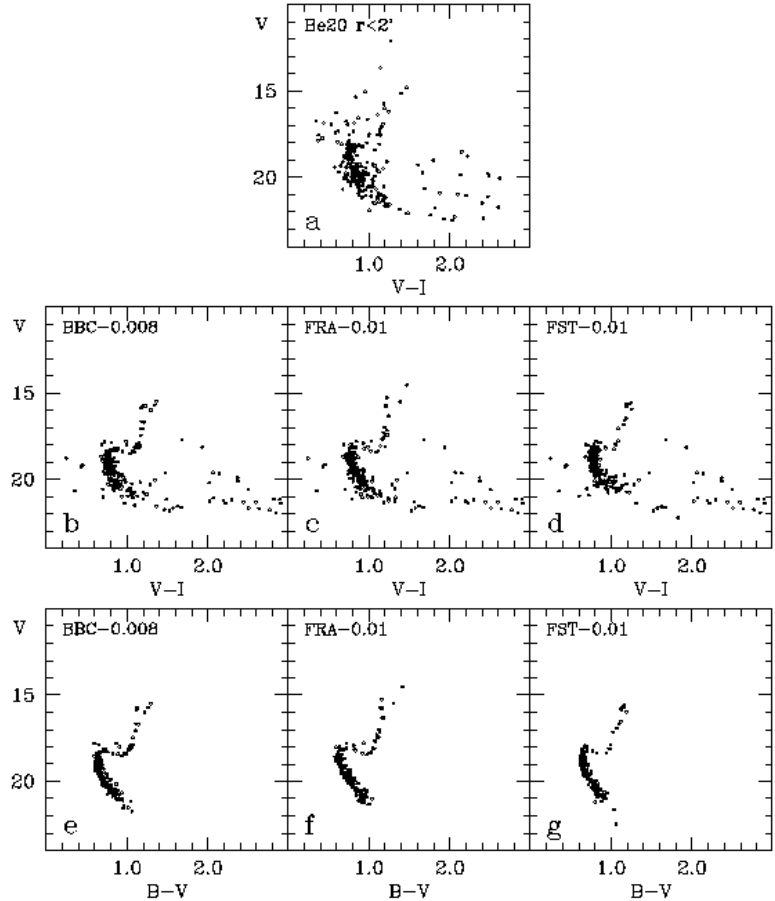} 
\caption{Comparison between observational and synthetic CMDs for Be~20. Panel
a  shows the stars measured in $V$, and $I$ in the central 2\arcmin ~radius
region.  Panels b, c, and d show the synthetic CMDs of best-fitting cases of
each type of stellar model, overimposed to the CMD of the same area in the 
control field for a more direct comparison. Panels e, f, and g show the 
corresponding $B-V$ CMDs. The displayed cases assume: Z=0.008, age=5.8 Gyr,
(m-M)$_0$=14.7, E$(B-V)$=0.13 for BBC, Z=0.01, age=4.3 Gyr,
(m-M)$_0$=14.7, E$(B-V)$=0.16 for FRA, and Z=0.01, age=5.0 Gyr,
(m-M)$_0$=14.7, E$(B-V)$=0.14 for FST.}
\label{sim20}
\end{figure*}

To minimize contamination without losing too many stars, we considered  as
reference CMD the diagram of the stars located within 2\arcmin\, from the 
cluster centre. As shown in Fig.~\ref{fig-radbe20}, this is the region better
compromising the two needs of removing contamination and having enough objects
to define the key evolutionary phases.  The CMD of this central circle is shown
in the top panel of Fig.~\ref{sim20}. It contains 269 stars, with 9 secure
members (from the RV) nicely defining the red clump and the RGB, as shown in the
central panel of Fig.~\ref{fig-cmdbe20}. Since the control field of the same 
area contains 70 stars, we assumed that the cluster members within  2\arcmin\,
are 200 and created the synthetic CMDs with this number of objects.

Fig.~\ref{sim20} shows the synthetic CMDs in better agreement with the data 
for each set of tracks, selected by minimizing the
differences between morphology, star counts, luminosity and colour functions in
the observational and synthetic MS, MSTO, SGB, RGB and clump phases. The top
panel of the Figure shows the reference CMD, the panels in the central row 
show the best synthetic $V, V-I$ CMDs obtained with the BBC, FRA, and FST models,
overimposed to the CMD of the control field stars. The $V, B-V$ synthetic CMDs
corresponding to the same cases are shown in the bottom panels.

\begin{figure}
\centering
\includegraphics[scale=0.65]{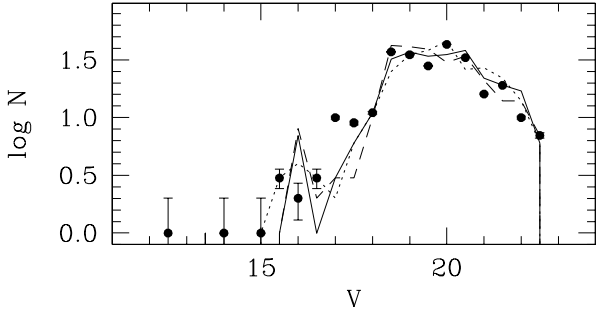} 
\caption{Comparison between observational (dots) and synthetic (curves) LFs 
of the stars within 2\arcmin\, from the centre of Be~20. The curves 
include the 200 synthetic stars and the 70 stars of the control field, and
correspond to the cases shown in Fig.~\ref{sim20}: solid for the BBC models,
dotted for the FRA and dashed for the FST ones. }
\label{simlf20}
\end{figure}

As apparent from Fig.~\ref{fig-radbe20}, field contamination affects mostly
the CMD portion fainter than $V\simeq$18, but interlopers are present also at
brighter magnitudes. Moreover, from that Figure and from the comparison of the
top panel with the central row panels of Fig.~\ref{sim20}, the CMD of the 
control field stars turns out to be not quite the same as that of the 
 cluster field contaminants. 
This circumstance, as well as the small number statistics,  weakens  the
discrimination power of the LF comparison. In fact, all the reasonable  cases
present LFs similar to each other, such as those displayed in 
Fig.~\ref{simlf20}. In this plot the dots show the LF of the 269 stars within
2\arcmin\, from the centre of Be~20, and the curves the LF of the  (synthetic +
field) stars of the CMDs in the central row panels of Fig.~\ref{sim20}. The
faint portions (dominated by field stars) of all the  curves are in good
agreement with the empirical LF, while the bright portions are consistent with
it, but generally underestimated. We ascribe this underestimation of the number
of relatively bright stars both to luminous interlopers and to the 
possible
presence of BSS, not accounted for in the  synthetic CMDs. 

\begin{figure}
\centering
\includegraphics[scale=0.99]{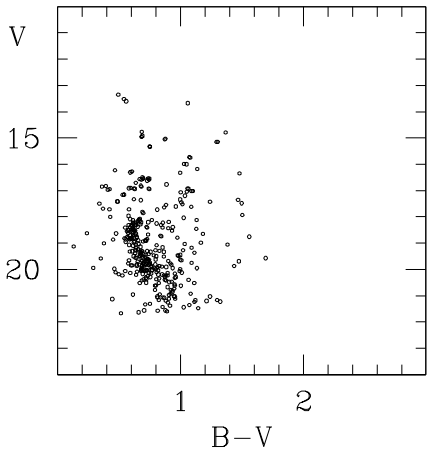} 
\caption{$V, B-V$ CMD of stars within 2\arcmin\, from the centre of Be~20, from
 Durgapal et al (2001).}
\label{bv20}
\end{figure}

Our photometry is only in two bands and, by itself, 
does not allow a reliable identification 
of the cluster metallicity. Models with solar metallicity or Z=0.004 do not
reproduce the CMD morphology as well as models with intermediate metallicity,
but cannot be excluded. Since the simultaneous agreement of the
synthetic $B-V$ and $V-I$ with the corresponding observational colours is a 
powerful metallicity indicator (see e.g. \citealt{tosi07} for King 11), we have
overcome our deficiency of information 
comparing the synthetic $V, B-V$ diagrams with the corresponding CMD published 
by \cite{d01}. The self-consistency of this solution is guaranteed by the 
agreement between our photometry and theirs, shown in Fig.~\ref{fig-conf1}. 
The $V, B-V$ CMD of the stars from \cite{d01} located within 2\arcmin\,  
from the cluster centre is shown in Fig.~\ref{bv20}. 
 
By comparing the synthetic diagrams based on all the stellar evolution sets and
on various assumptions on the age, distance and reddening of Be~20 with our $V,
V-I$ CMD and  with the $V, B-V$ CMD from \cite{d01}, we are able to distinguish
quite well all the cluster parameters. We find that only models with Z=0.008 or
0.01 can simultaneously reproduce the $B-V$ and $V-I$ observed colours. For
higher metallicities $B-V$ is always too red when $V-I$ is fine, and, vice
versa, for lower metallicities $B-V$ is always too blue when $V-I$ is fine.
 Hence, we assign to Be~20 a metallicity Z=0.009$\pm$0.001. 
This  photometric estimate of the cluster metallicity is in perfect agreement
with  that inferred by \cite{sestito08} from high-resolution spectroscopy 
([Fe/H]=$-0.3$). 

The age providing the best reproduction of the observed CMD morphology and
stellar density in the different evolutionary phases depends on the assumptions
of the stellar evolution tracks and is therefore slightly different from one set
of models to the other. In spite of the field contamination and small number
statistics, the clear identification from RVs of the SGB and clump stars allows
us to infer the cluster age quite strictly. With the FRA models the age is in
the small range 4.3 -- 4.5 Gyr: older cases have the clump too bright, and
younger cases have it too faint. With overshooting models, the age is obviously
older, but only by less than 25\% because in these low mass stars overshooting
is not very effective. With the FST models the best-fitting ages are between 5
and 5.5 Gyr, while with the BBC models they are between 5.5 and 5.8 Gyr.

Reddening and distance modulus are also well identified by best-fitting
magnitudes and colours of the main evolutionary phases. Particularly striking is
the circumstance that  all the models in agreement with the data require
(m-M)$_0$=14.7. E$(B-V)$ is more sensitive to the metallicity and the details of
the stellar models, and we find it to range between 0.13 (for all the BBC-0.008 
good cases) and 0.16 (corresponding to the FRA-0.01 case with age 4.3 Gyr).

In summary, thanks to the cluster membership provided by the radial velocities, 
age, distance, reddenning and metallicity of Be~20 are very well identified. 


\subsection{Be~66} \label{param_be66}

The situation for Be~66 is apparently more complicated, both because of its 
high extinction and of the lack of information on membership. The fraction of
fore/background contaminating objects doesn't significantly vary from the
cluster centre to the periphery and we have therefore chosen to simulate the
whole CMD. Since the stars measured  in all the three $B, V$ and $I$ bands are
2362 in our cluster field and 1023 in the control field, the synthetic CMDs
have been created with 1339 objects, assuming the photometric errors and the
completeness factors described in Sect.~\ref{data}.

\begin{figure*}
\centering
\includegraphics[scale=0.65]{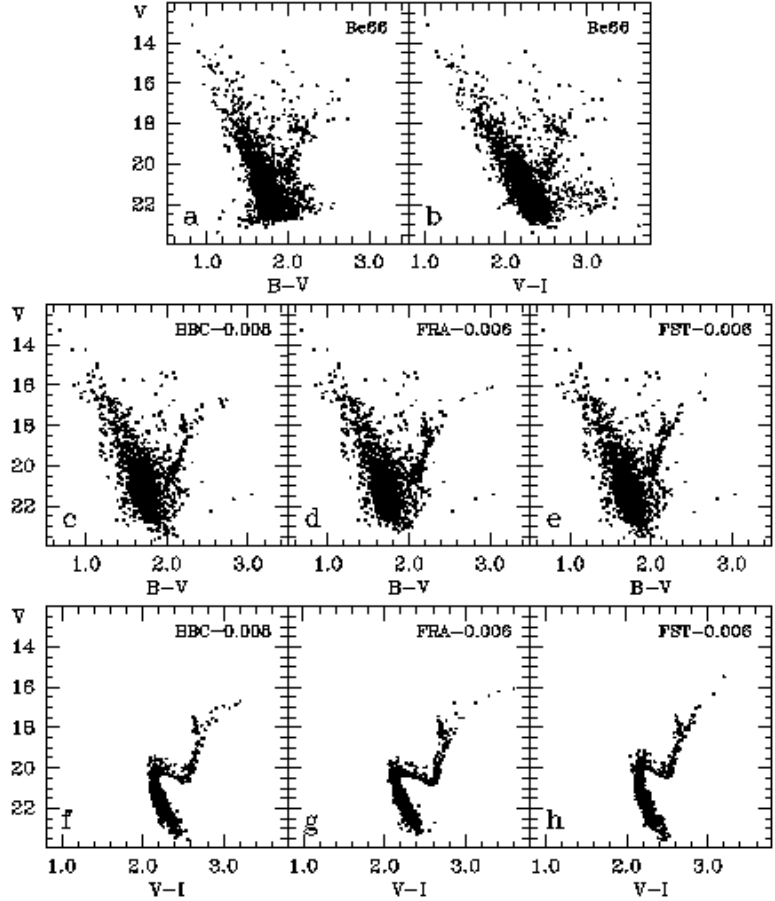} 
\caption{Comparison between observational and synthetic CMDs for Be~66. Panels
a and b show the stars measured in $B$, $V$, and $I$ in our cluster field.  
Panels c, d and e show the synthetic $V, B-V$ CMDs of best-fitting cases of
each type of stellar model, overimposed to the CMD of the same area in the 
control field for a more direct comparison. Panels f, g and h show the 
corresponding $V, V-I$ CMDs. The displayed cases assume: Z=0.008, age=3.8 Gyr,
(m-M)$_0$=13.3, E$(B-V)$=1.20, 1.22, 1.24 for BBC; Z=0.006, age=2.7 Gyr,
(m-M)$_0$=13.5, E$(B-V)$=1.28, 1.30, 1.32 for FRA, and Z=0.006, age=3.0 Gyr,
(m-M)$_0$=13.3, E$(B-V)$=1.24, 1.26, 1.28 for FST.}
\label{sim66}
\end{figure*}

The major problem encountered by all models is the width of the evolutionary 
sequences, both in colour and in magnitude. In no way are we able to create
synthetic diagrams with the observed width by including only  photometric errors
and binary systems. However, the synthetic sequences become properly thick if we
assume variable amounts of reddening. We find that a $\delta$\ebv=$\pm$0.02 is
sufficient to account for the observed spread. Given the high reddening
affecting Be~66, this variation corresponds to only 2-3 per cent and looks
rather likely.

The widths of the MSTO, SGB and clump could also make it more difficult to 
precisely define the cluster parameters. However, it turns out that the
independent constraints from the colour, magnitude, morphology, stellar density
of the different evolutionary sequences and luminosity function do allow  to
significantly reduce the range of possible values. The distance modulus, for
instance, turns out to be strikingly stable in all the models in acceptable
agreement with the data: in spite of its dependence on age, reddening and
metallicity, we find $13.2\leq$ \mmm $\leq$13.5 in all the acceptable cases,
and \mmm=13.3 in all the best cases.

To identify the cluster metallicity, the clue is the simultaneous consistency 
of both the predicted \bv~ and \vi~ colours with the observed ones.  All the
models with solar metallicity and right \bv~ predict  too blue \vi, so we can
therefore exclude Z=0.02. For lower metallicity, the different sets of stellar
tracks have the following responses: at Z=0.01, the FST models still show the
\bv~ inconsistency with \vi~  of the solar models, although at a lower level,
while the FRA models cannot be excluded.  The BBC models with Z=0.008 have
self-consistent colours and acceptable CMD and LF properties. The FST models
with Z=0.006 perfectly match the observational  CMD and LF, and the FRA models
with Z=0.006 also have self-consistent  colours and accetable CMDs and LFs. Also
the BBC models with Z=0.004 give self-consistent colours and acceptable CMDs and
LFs.  Within the uncertainties, all these results indicate that the metallicity
of Be~66 is most likely  Z=0.006$\pm$0.002. This value corresponds to
[Fe/H]$\simeq-0.45$, lower than the range proposed by \cite{pj96} but in perfect
agreement with the clump star abundance measured by \cite{villanovabe66} from high
resolution spectroscopy.

The reddening resulting from the synthetic CMDs depends both on the adopted
models and metallicities. In the cases in better agreement with the data, we
find it to range between \ebv $=1.22\pm0.02$ and $1.30\pm0.02$, where 0.02 is
the variation necessary to reproduce the observed spread of the evolutionary
sequences. In the best-fitting case (FST with Z=0.006 and $\tau$=3 Gyr)
\ebv=$1.26\pm0.02$.

The age depends on the various assumptions as well. However, in spite of the
width of the evolutionary sequences and the uncertainties on reddening and
metallicity, we find it to be rather well determined for each kind of stellar
models, thanks to the well defined shape and stellar density of key phases, such
as MSTO, SGB, RGB and clump. With the FRA models without overshooting we find
$2.7\leq\tau/Gyr\leq$3.0, both for Z=0.006 and
Z=0.01, with the preferred value between 2.7 and 2.8 Gyr. The models with
overshooting provide $3.5\leq\tau/Gyr\leq4.0$ with the BBC models with either
Z=0.004 and Z=0.008 (preferred value 3.8 Gyr), and $3.0\leq\tau/Gyr\leq3.5$ 
with the FST models with intermediate overshooting (preferred value 3.0 Gyr).
 
\begin{figure}
\centering
\includegraphics[scale=0.65]{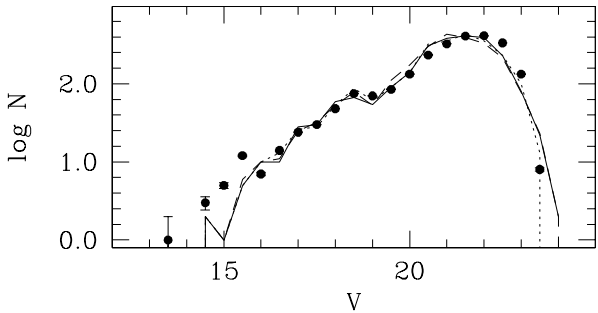} 
\caption{Comparison between observational (dots) and synthetic (curves) LFs 
of the stars  in Be~66. The curves 
include the 1339 synthetic stars and the 1023 stars of the control field, and
correspond to the cases shown in Fig.~\ref{sim66}: solid for the BBC models,
dotted for the FRA and dashed for the FST ones.}
\label{simlf66}
\end{figure}

Fig.\ref{sim66} shows the synthetic CMDs in better agreement with the $V, B-V$ 
and $V, V-I$ diagrams for each set of tracks, selected by minimizing the
differences between morphology, star counts, luminosity and colour functions in
the observational and synthetic MS, MSTO, SGB, RGB and clump phases. The two top
panels of the Figure show the observational CMDs, the bottom panels show the
best synthetic $V, V-I$ CMDs obtained with the BBC, FRA and FST models, while
the panels in the central row show the $V, B-V$ CMDs of the same cases as in the
bottom row, overimposed to the CMD of the control field stars.

Fig.\ref{simlf66} compares the cluster LF with the LFs from the synthetic 
cases  shown in Fig.~\ref{sim66}. The three models predict LFs so similar to
each other that the three curves overlap almost completely. They all agree very
well with the data, except at the bright end, where field contamination
dominates and the difference between the foreground stars in the cluster field
and in the control field is apparent.

For Be~66, the FST models with Z= 0.006, age=3 Gyr, \ebv=1.26$\pm$0.02 and
\mmm=13.3 are by far the ones in better agreement with the data. 

\subsection{To~2} \label{param_to2}

\begin{figure*}
\centering
\includegraphics[scale=0.65]{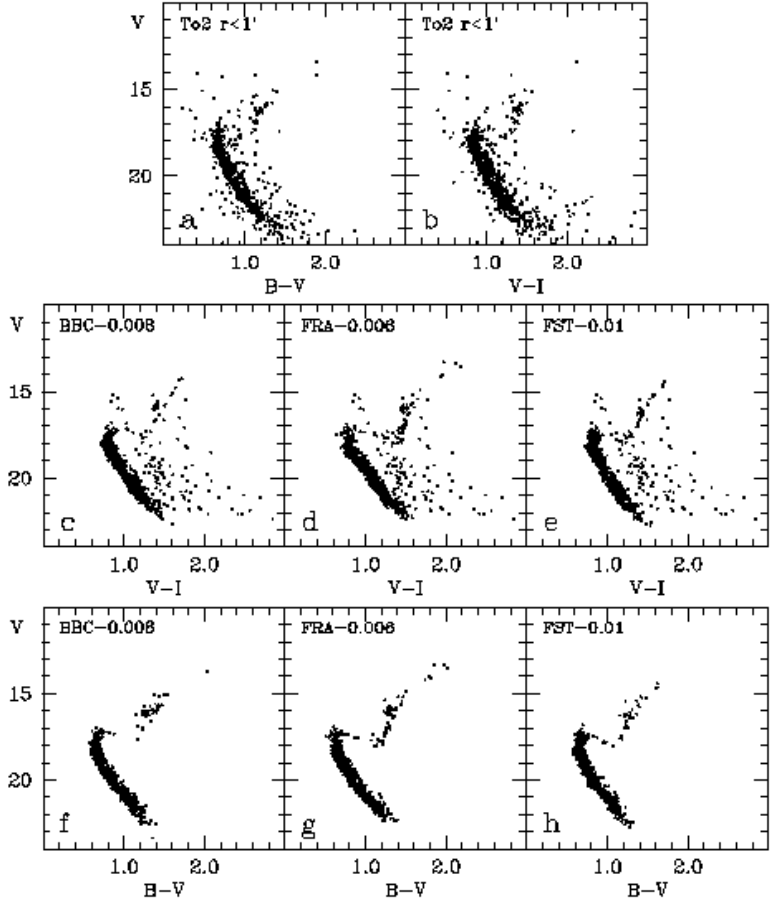} 
\caption{Comparison between observational and synthetic CMDs for To~2. Panels
a and b show the stars measured in $B$, $V$, and $I$ in the central 1\arcmin 
~radius region.  Panels c, d, and e show the synthetic CMDs of best-fitting cases of
each type of stellar model, overimposed to the CMD of the same area in the 
control field for a more direct comparison. Panels f, g, and h show only the
synthetic stars of the 
corresponding $B-V$ CMDs. The displayed cases assume: Z=0.008, age=1.6 Gyr,
(m-M)$_0$=14.5, E$(B-V)$=0.34 for BBC, Z=0.006, age=1.4 Gyr,
(m-M)$_0$=14.5, E$(B-V)$=0.0.40 for FRA, and Z=0.01, age=1.7 Gyr,
(m-M)$_0$=14.7, E$(B-V)$=0.31 for FST.}
\label{simto2}
\end{figure*}

As discussed in the previous Sections, To~2 presents a fairly complicated CMDs,
strongly affected by (probably multiple) contamination. We did observe a nearby
region as control field to evaluate the contamination, but the comparison of its
CMD with that of To~2 clearly shows that the two populations are rather different
from each other (see Fig.\ref{fig-cmdto2}). Moreover, we didn't have the opportunity 
to acquire images of
the external field in all the filters, and B is missing. Hence, the control 
field is of little help to decontaminate To~2. 
From the morphology of To~2's 
CMDs with increasing distance from the cluster centre it is apparent that in the
circle within 2 arcmin the contamination is still rather significant (at least 15\%,
according to the number of stars in an equal area region of the control field).
To minimize it, we thus decided to adopt as reference cluster CMD that of the
central region of 1 arcmin radius (see Fig.\ref{fig-binto2}). It contains 1147 objects with measured  V
and I magnitudes, and 846 with B, V and I. Since control field portions of 
equal size contain 100 stars with measured V and I, we can assume a 9 percent
(lower limit to) contamination. In other words, of the 846 objects with measured
BVI, 770 can be considered To~2 members. The synthetic BVI CMDs have been
therefore computed assuming this number of objects.
The B-V and V-I CMDs of the 846 BVI objects in the central region of To~2 are
displayed in the top panels of Fig.\ref{simto2}.

Also in this case, we have tried to blindly identify the metallicity, without biassing our
search on the basis of literature estimates, for homogeneity with the BOCCE project procedure. 
All the models clearly exclude a solar metallicity: whatever the adopted 
tracks, Z=0.02 always lead to CMDs inconsistent with the data both
because of the MS shape and of the colours. If we assume a reddening allowing to
reproduce the B-V colours, then V-I is systematically too blue, and vice versa.
 Moreover, the
acceptable reddenings are systematically lower than in literature.
While rejecting a solar metallicity is straightforward, finding the best fitting
one is extremely difficult, because contamination affects key features before
and after the MSTO. Since a good identification of the metallicity is necessary
to identify the reddening, our analysis leaves the latter rather uncertain too.
Vice versa, the distance modulus of To~2 turns out only moderately sensible to
the metallicity choice and between 14.4 and 14.7 for any viable model.

With the BBC models, the available metallicity most appropriate
for the CMD of To~2 is Z=0.008, whose models allow to self-consistently
reproduce the observed B-V and V-I colours of all the evolutionary phases.
However, the corresponding upper MS and MSTO regions never have exactly the same
morphology as the observed ones. The best CMD with these tracks assumes  age
$\tau$ = 1.6 Gyr, reddening E(B-V) = 0.34 and distance modulus (m-M)$_0$=14.5.
Its CMDs are plotted in panels c and d of Fig.\ref{simto2}. Panel f shows
only the 770 synthetic stars, while panel c contains 846 objects like the
empirical CMDs of the top panels, since it includes also the 75 objects falling
in an equal area portion of the control field. The comparison of panels b and 
c emphasizes the difference between the control field and the cluster
contaminating stars. 
BBC models with Z=0.004 also lead to an acceptable self-consistency between 
B-V and V-I colours, but systematically have RGBs redder and MSs straighter 
than observed. For this reason we consider them less appropriate for To~2 than
the Z=0.008 ones.

For the other types of models, both the FRA and the FST metallicities below
solar are Z=0.01 and Z=0.006, and for both it is impossible to significantly
discriminate between them. With the FRA models, both metallicities allow for
self-consistent colours, but both provide RGB always quite redder than observed.
To shrink the subgiant branch and keep the RGB within the observed colour range,
one should increase the age, but this inevitably implies clumps brighter than
observed. Generally speaking none of the FRA models reproduces adequately the
properties of To~2's CMD. In panels d and g of Fig.\ref{simto2} we simply show
the case in smaller disagreement, with no claim of actual consistency. It
corresponds to Z=0.006, age 1.4 Gyr, E(B-V)=0.40 and distance modulus
(m-M)$_0$=14.5.

With the FST models, discerning the right metallicity is equally difficult,
because Z=0.006 leads to the best self-consistency between B-V and V-I, but
predicts RGBs slightly redder than observed, while Z=0.01 has V-I slightly too
blue when B-V is correct (a typical signature of a metallicity overestimate),
but appropriate RGBs. At variance with the FRA models, the FST ones 
provide however a good agreement with the empirical CMD, actually better than with any
of the other models. Based on the overall properties of the CMD (curvature of the
MS, shape of the MSTO and subgiant branch, colour and slope of the RGB, morphology
of the clump, number of stars in each of these phases) we ended up preferring
the Z=0.01 models. Most likely, the right metallicity (at least within the
framework of the FST models) is slightly below this value.
In panels e and f we have chosen to show one of the best
cases among the FST ones: age 1.7 Gyr, E(B-V)=0.31 and distance modulus
(m-M)$_0$=14.7. Here the choice is rather subjective: an age of 1.6 Gyr (with
same metallicity and distance modulus and E(B-V)=0.32), or an age of 1.8 Gyr 
(with same metallicity, E(B-V)=0.30 and distance modulus (m-M)$_0$=14.5) would 
have been good as well.

Since the FST models come in three overshooting flavours (see Table 7), 
we have tested them all. As often found for the BOCCE clusters, the FST models with
overshooting reproduce the CMD of To~2 much better than those without. 
In this case, we found no significant difference in the quality of the fit from
models with the highest or intermediate overshooting.

We have tested various fractions of unresolved binaries and found that while
their presence is clearly required to reproduce the secondary MS seen 
(see Fig.\ref{fig-binto2}) at the
right of the main MS both in B-V and in V-I (hence a binary fraction of 0 must 
be rejected), we don't need to invoke particularly high percentages. A 60\%
fraction would definitely lead to an excessive blend of the binary and single 
star MSs and is thus beyond the acceptable value. We conclude that a fraction 
around 30\% provides a good representation of the data.

\begin{figure}
\centering
\includegraphics[scale=0.65]{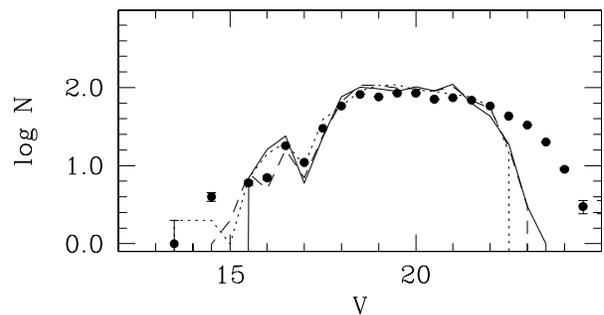} 
\caption{Comparison between observational (dots) and synthetic (curves) LFs 
of the stars within 1\arcmin\, from the centre of To~2. The curves 
include the 770 synthetic stars and the 76 stars of the control field, and
correspond to the cases shown in Fig.~\ref{simto2}: solid for the BBC models,
dotted for the FRA and dashed for the FST ones. }
\label{simlfto2}
\end{figure}

The difficulty in properly characterizing the field contamination has the
further consequence of reducing the selection power of the comparison between
synthetic and empirical LF. In Fig.\ref{simlfto2} the LFs of the three models
plotted in panels c, d and e of Fig.\ref{simto2} are plotted as lines, while the
empirical LF of the 846 stars of the central 1\arcmin of To~2 is represented by
dots with error bars. The synthetic LFs fit reasonably well the portions where
the cluster dominates, but at both the faint and the bright ends, where
contamination dominates and the control field is not representative of that
population, all the LFs fail, inevitably unable to predict the number of
interlopers. Notice however that in the central parts of the plot, the
agreement is fairly good. In particular the bump and dip corresponding to the
clump are well fitted, showing that both age and distance modulus are 
appropriate.

To summarize, we conclude that within the high uncertainties, To~2 turns out to
be a cluster of intermediate age (1.4 Gyr for models without overshooting,
1.6-1.8 Gyr for models with overshooting), with distance modulus 14.5-14.7, and
with reddening between 0.30 and 0.40 depending on the chosen metallicity, but
most likely within 0.31-34 (i.e. for 0.008$\leq$Z$\leq$0.01).

It is interesting to notice that, in spite of the uncertainty in the
identification of the best metallicity, we don't find any need
of a metallicity spread as found by \cite{frinchaboy08}. Assuming a metallicity or
reddening dispersion in our synthetic CMDs would introduce a spread in the
evolutionary sequences larger than observed. 
Notice that only their metal-poor population falls in the central 1\arcmin-radius circle, so
our result  of a single population of
stars with a single metallicity 0.008$\leq$Z$\leq$0.01 is in perfect agreement
with theirs, as are also reddening and distance modulus.
Since in the CMD of the region with 2\arcmin-radius we already see significant
contamination, and since the most central metal-rich star is at least
1.5\arcmin from To2's centre (see their Fig. 2), of the various alternatives 
examined by Frinchaboy et al. we
favour the interpretation of the metal richer populations in terms of an
interloper.
The more recent finding of absence of any metallicity dispersion \citep{villanovato2}
is even more easily in agreement with our study. The metallicity is similar, as are the 
distance, reddening, and age. 

\begin{table}
\centering
\caption{Summary of the parameters for the three clusters. We use the BBC results, for homogeneity in ranking and immediate compatibility with the 16 OCs presented in Table 1 of Bragaglia \& Tosi (2006); see text for the values derived using the other tracks. Note that Be~66 has differential reddening.}
\setlength{\tabcolsep}{1.6mm}
\begin{tabular}{lcccccc}
\hline\hline
OC  & age & $(m-M)_0$ & d$_\odot$ & $R_{GC}$ & $E(B-V)$ & Z \\
        &(Gyr)&                      &(kpc)            &(kpc)           &                  &     \\
\hline
Be~20 & 5.8 & 14.7 & 8.71 & 16.0 & 0.13 & 0.008\\
Be~66 & 3.8 & 13.3 & 4.57 & 12.0 & 1.22 & 0.008\\
To~2   & 1.6 & 14.5 & 7.95 & 14.3 & 0.34 & 0.008 \\
\hline
\end{tabular}
\label{tab-param}
\end{table}

\section{Summary and discussion} \label{discussion}

We compared the CMDs of the three old, distant OCs Be~20, Be~66, and To~2 to synthetic
ones based on three different sets of evolutionary tracks and determined the
clusters parameters. Table~\ref{tab-param} gives a summary of the derived parameters for the three OCs. We use here the values based on the BBC tracks, even when they do not offer the best-fitting solution (e.g., for Be~66), in analogy to what we did in \cite{bt06} to obtain a homogeneous ranking on a single scale. 
\begin{itemize}
\item All three clusters are less metal-rich than the Sun, with best=fitting solutions of
Z=0.008-0.01 (depending on the set of tracks) for Be~20 and To~2, and Z=0.006 for Be~66.
These rather precise values could be obtained because we considered the
simultaneous good fit of both the $V-I$ and $B-V$ CMDs.
\item
The reddenings slightly depend on the
tracks metallicity.  For Be~66, we confirm \cite{pj96}'s
suggestion of a probable differential reddening, of the order of 2-3 per cent, a
rather likely occurrence given the high extinction level.
\item
For all clusters a binary fraction of about 30 per cent seems necessary 
to well reproduce the width of the observed sequences.
\item
The derived ages depend on the treatment of convection adopted in the
evolutionary tracks, with the usual lower values found for tracks without
overshooting.  
\item Our results on age, distance and reddening do not significantly differ from the
most recently published ones for Be~20 and To~2.
We find Be~66 slightly younger, closer, and definitely metal poorer than
\cite{pj96}, while we agree with them on the reddening and with 
\cite{villanovabe66} on the rather low metallicity.
\item As already anticipated in Sect.~\ref{intro}, these OCs have Galactocentric radii of about 12 to 16 kpc and are then useful to constrain the properties of the outer disc, in particular in what seems to be a transition region for the metallicity distribution (see below).
\end{itemize}

\begin{figure}
\centering
\includegraphics[bb=75 180 550 680, clip, scale=0.45]{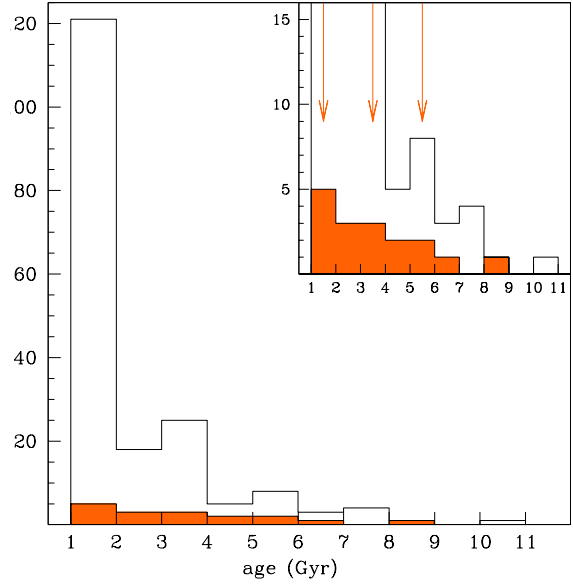}
\caption{ Age distribution for the 186 OCs older than 1 Gyr, according to the 2010 update of
the Dias et al. (2002) catalogue. The
filled histogram represents the 17 clusters in the BOCCE sample within the same age limit (14 already published, three in the present paper).
The inset shows an enlargement to better appreciate
the -still small- number of very old objects, and the three new OCs are indicated by arrows.}
\label{fig-old}
\end{figure}

\noindent{\bf The Galactocentric metallicity gradient -} 
Even if the number of known
old OCs has steadily grown in the last years, they still are a minority, since
the \cite{dias02}  catalogue contains about 2000 objects,
and only about 190 are older than 1 Gyr (see Fig.~\ref{fig-old}). An even smaller number of clusters
has the metallicity determined using high-resolution spectroscopy (see below). In
the BOCCE project we have especially targeted old clusters, hence our OCs
represent a fair sample of the old cluster population, about 10 per cent.  In  Fig.~\ref{fig-old} 
we also show for comparison the same histogram for our 
sample,\footnote{This comparison is only  indicative, since age determinations
are not homogeneous between our work and the catalogue (and inside the catalogue
itself). There are differences, sometimes large, between ages measured by
different authors. For instance, in \cite{dias02}, a)  one of the two clusters in
the last bin  -age 10 to 11 Gyr- is Be~17 that, according to \cite{bra06a} is
instead less than 9 Gyr old, b) King~11  is given an age of 1.1 Gyr, while we
derived an age  $\gsim$ 4 Gyr \citep{tosi07}, in line with other determinations,
and c) NGC~6791 is  attributed an age, 4.4 Gyr, younger than found in all
recent papers, where its age is 7-9 Gyr  \citep[see, e.g,][]{king05}.  }
and indicate the three clusters analysed here in the enlargement.

The metallicity distribution in the disc is an important  ingredient of chemical
evolutionary models, and offers essential information on the formation and
evolution of the disc, especially when also its possible evolution with time is
considered. We may obtain the present-day metallicity  distribution from O,B
stars and H {\sc ii} regions \citep[see, e.g.,][]{rudolph}.  Another possibility
are Cepheids
\citep[e.g.,][]{andrievsky}, for which the distance and age can 
be determined with high precision.
Planetary nebulae (PNe) are another tracer, and they are in
principle able to reach further in the past \citep[see, e.g., the recent paper
by][]{sh}; however, there still are difficulties in assigning individual ages
and distances.  This is in general true for all field stars; the most accurate
parallaxes to date, those measured by the Hipparcos space mission, are available 
only for the Sun's vicinity. Ground-based catalogues can reach farther away, but at
the price of accuracy. HST parallaxes can be very precise, but only a handful of targets has been observed. To obtain
precise distances
for hundred millions stars in the whole Galaxy we will have to wait about 10 years, for the 
completion of the Gaia\footnote{See http://www.rssd.esa.int/index.php?project=GAIA\&page=index.}
satellite survey. 

Furthermore, all stars are subject to orbit migration \citep[see,
e.g.,][in the recent years]{sellwood,roskar,sb09};  they move away from their
birthplace even several kpc, thus complicating the study of the metallicity
distribution: at any given $R_{GC}$ we may find stars born there, or in an inner
(i.e., in general more metal-rich) or in an outer (i.e., in general more
metal-poor) region of the disc. 

OCs are less subject to all these problems: their distances and ages can be
measured with sufficient precision using the stellar models \citep[e.g., as we
do in our BOCCE project, see][]{bt06}, their abundances can be determined from
several or many member stars \citep[e.g.,][]{bragaglia08}, and they do not
appear to suffer from orbit migration.  For the last point, see e.g., \cite{wu},
who computed the orbits of about 500 OCs, using information from \cite{dias02}.
When they derive the slope of the metallicity gradient using $R_{GC}$ or the
apogalacticon radius, they do not find any significant difference. Apart from a
few exceptions, it appears safe to use OCs and their present-day positions, to
define the metallicity distribution now and in the past. This is part of our goals,
and we are building the BOCCE sample choosing OCs which cover the whole distribution
of clusters' properties. Of course, the
situation will further improve when it will be possible to compute orbits for
the whole family of OCs, using the improved distances, proper motions, and
radial velocities produced by Gaia.

\begin{figure}
\centering
\includegraphics[bb=40 200 550 680, clip, scale=0.47]{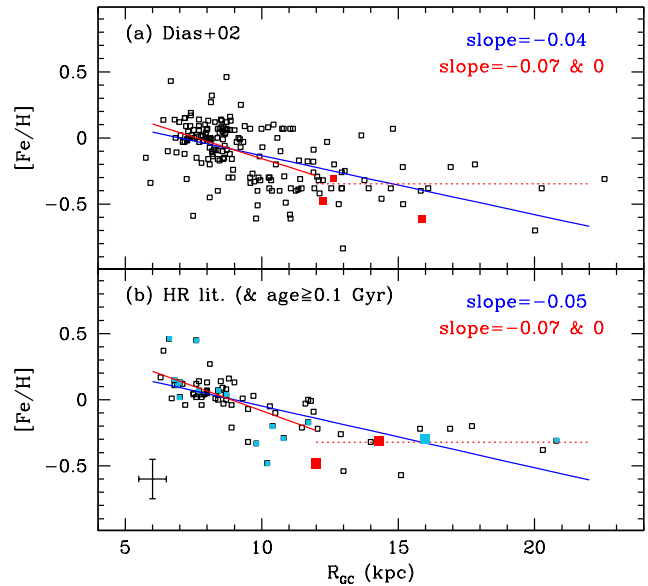}
\caption{ The metallcity gradient as defined by OCs. 
(a) R$_{GC}$ and [Fe/H] are taken from Dias et al. (2002). The two lines represent the fit to the data using a single slope 
(in blue) or two, within 12 kpc from the Galactic centre and farther than this limit (in red), see text. 
The three clusters of the present paper are indicated by filled (red) squares. 
(b)The same, but using only 72 OCs older than 0.1 Gyr and for which the metallicity has been derived with high-resolution spectroscopy (the three clusters are shown with larger
symbols); filled, light-blue squares indicate OCs for which [Fe/H] has been determined by our group.}
\label{grad1}
\end{figure}

We show in Fig.~\ref{grad1}(a) the radial metallicity gradient described by OCs
using the information in the \cite{dias02} catalogue. We computed $R_{GC}$ using
8 kpc as the Sun distance from the Galactic centre, and the tabulated distances
from the Sun. Values for [Fe/H]  come from photometry (CMD, narrow-band,
indices), low resolution, and high resolution spectroscopy; individual
references  can be found at the Dias webpage
(http://www.astro.iag.usp.br/$\sim$wilton/). We display with filled (red) symbols the postion of Be~20, Be~66, and To~2, using the tabulated values. The metallicity tends to decline
from the centre to the outskirts of the disc. If we fit the distribution with a
single relation, we find a decline rate of $-0.04~{\rm dex~kpc}^{-1}$, as indicated
in the figure. However, recent studies of open clusters (e.g.,
\citealt{yong05,carraro07,sestito08,friel10}) suggest that a better
representation is a  gradient in the inner region ($\le$12 kpc from the
centre) followed by an almost flat value therefore. We then divided the sample of OCs accordingly, computing a (steeper) inner slope
of $-0.07~{\rm dex~kpc}^{-1}$. The OCs in the external part have in this case
an average metallicity of =-0.35 dex.
While the number of OCs in this sample is large (177 objects), they tend to lie mostly in the
Sun's vicinity, with only about 15 per cent of them with $R_{GC}>12$ kpc, and only a handful 
in the outermost disc regions. Furthermore, the sample is completely inhomogeneous.

In principle, choosing only [Fe/H] measured using high resolution spectroscopy
produces more solid results. We have searched the literature and retrieved the
metallicity, distance, and age of about 70 clusters. A good fraction of them, about 30 per cent,
are also in the BOCCE sample and are indicated with filled symbols in the figure; we show the three clusters 
of the present paper with larger symbols, this time using the parameters derived in the present paper and the [Fe/H] values indicated in Table~\ref{deltav}. We restricted
our search to clusters older than 0.1 Gyr and selected in each case either the
most recent determination or our measures, when present. The resulting
metallicity distribution is shown in  Fig.~\ref{grad1}(b). The slopes derived
are (almost) the same as in the previous case. In this case, however, there is
a better balance between inner and outer clusters, and the individual values are
more precise (but again, with a caveat on the inhomogeneity of sources, which
produces systematics).  Be~66, To~2, and Be~20, in order of increasing distance from the centre,
are an important addition, since they are about one fifth of the entire outer-disc sample.

\begin{figure}
\centering
\includegraphics[bb=40 170 550 650, clip, scale=0.47]{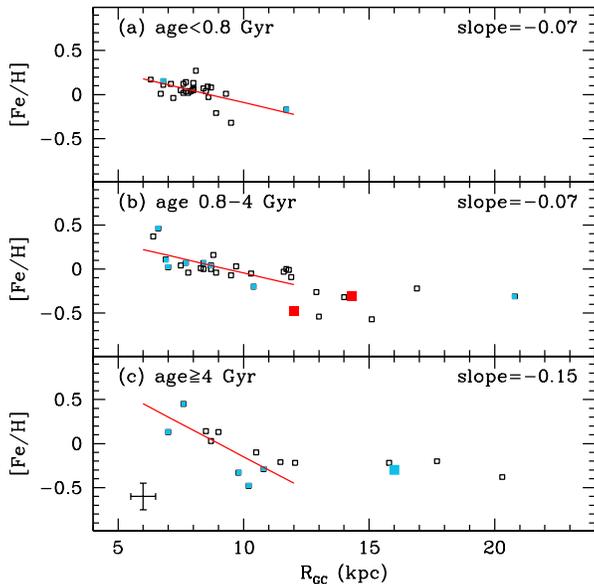}
\caption{The Galactocentric metallicity distribution, in three different time
intervals (symbols are as in the previous figure). The slopes, computed only for the inner 12 kpc, are indicated.}
\label{grad2}
\end{figure}

This second sample of OCs can also be used to study the behaviour of the gradient with
time. In Fig.~\ref{grad2} we divide the sample in three parts and compute the
slope of the gradients (only in the inner 12 kpc) again, with the three clusters hightlighted. 
While young and intermediate
age clusters seem to share the same behaviour, the slope of the gradient seems
to have been steeper in the past (as found also for PNe).  This has to be
accounted for by any chemical evolution model. Note however the paucity of
very old clusters at the transition between a decreasing and flat distribution of
metallicities; this calls for new additions to the well-studied cluster sample.

We plan to repeat these exercises using only clusters in the BOCCE sample, with
ages, distances, and [Fe/H] all derived on the same scale. The importance of
Be~20, Be~66, and To~2 lies both in their old ages and in  their large $R_{GC}$
 (see e.g., the sample map in \citealt{b08}).  Be~66 is the outermost
cluster in the second quadrant,  while Be~20 and To~2 fill the large gap in
$R_{GC}$ between Be~29 (at 20.8 kpc) and Be~22 (at 14.0 kpc). Notice that in the
literature very few clusters have been studied beyond a radius of 16 kpc. For
instance, \cite{carraro07} list only Be~31, Be~73, Be~25, Saurer~1, and Be~29. 

Different interpretations of the two-slopes metallicity  distribution are
possible, either a normal outcome of disc formation and chemical enrichment
\citep[e.g.,][]{carraro07,sb09} or  satellite accretion in the outer disc (e.g.,
\citealt{yong05}). The former suggestion likely foresees a continuous radial
distribution (even if with different slopes, maybe even zero, in different regions), while the latter implies some inhomogeneity reflecting the
satellite impact trajectory. It is then important to have as  many clusters as
possible outside the critical radius of 12 kpc, and to have them in different
quadrants.

We  will use the information obtained here, together with the detailed chemical
abundances from existing and future high resolution spectroscopic data, to
increase the BOCCE set and derive conclusions based on a homogeneous analysis.

\section*{Acknowledgements}

We gratefully acknowledge the use of software written by P. Montegriffo,
and of the  WEBDA database, created by J.C. Mermilliod,
and now operated by E. Paunzen at the Institute for  Astronomy of the University of Vienna.
We thank Mara Manduchi for her work on To~2.
This work has made use of  VizieR, operated at CDS, Strasbourg, 
France, and of NASA's Astrophysics Data System.
AB acknowledges the hospitality of ESO Chile (through the Scientific Visitor
Programme) where part of this work was done.


\begin{thebibliography}{99}

 \bibitem[\protect\citeauthoryear{Adler \& Janes}{1982}]{adler_janes} 
 Adler D.~S., Janes K.~A., 1982, PASP, 94, 905 

 \bibitem[\protect\citeauthoryear{Andreuzzi et al.}{2004}]{andreuzzi} 
 Andreuzzi G., Bragaglia A., Tosi M., Marconi G., 2004, MNRAS, 348, 297 

 \bibitem[\protect\citeauthoryear{Andrievsky et al.}{2004}]{andrievsky} 
 Andrievsky S.~M., Luck R.~E., Martin P., L{\'e}pine J.~R.~D., 2004, A\&A, 413, 159 

 \bibitem[\protect\citeauthoryear{Bellazzini et  al.}{2004}]{bellazzini04} 
 Bellazzini M., Ibata R., Monaco L., Martin N., Irwin M.~J., Lewis G.~F., 2004, MNRAS, 354, 1263 

 \bibitem[\protect\citeauthoryear{Bessell, Castelli \& Plez}{1998}]{bessel98} 
 Bessell M.~S., Castelli F., Plez B., 1998, A\&A, 333, 231 

 \bibitem[\protect\citeauthoryear{Bragaglia}{2007}]{b08} 
 Bragaglia A., 2007, arXiv, 711, arXiv:0711.2171  

 \bibitem[\protect\citeauthoryear{Bragaglia}{2010}]{bragaglia10} 
 Bragaglia A., 2010, IAUS, 268, 119 

 \bibitem[\protect\citeauthoryear{Bragaglia \& Tosi}{2006}]{bt06} 
 Bragaglia A., Tosi M. 2006, AJ, 131, 1544

 \bibitem[\protect\citeauthoryear{Bragaglia et al.}{2006a}]{bra06a}
 Bragaglia A., Tosi M., Andreuzzi G., Marconi G., 2006a, MNRAS, 368, 1971 
 
 \bibitem[\protect\citeauthoryear{Bragaglia et al.}{2006b}]{bra06b} 
 Bragaglia A., Tosi M., Carretta E., Gratton R.~G., Marconi G., Pompei E., 2006b, MNRAS, 366, 1493 

 \bibitem[\protect\citeauthoryear{Bragaglia et al.}{2008}]{bragaglia08} 
 Bragaglia A., Sestito P., Villanova S., Carretta E., Randich S., Tosi M., 2008, A\&A, 480, 79 
 
 \bibitem[\protect\citeauthoryear{Bressan et al.}{1993}]{Bres93} 
 Bressan A., Fagotto F., Bertelli G., Chiosi C., 1993, A\&AS, 100, 647 

\bibitem[\protect\citeauthoryear{Brown et al.}{1996}]{brown} 
 Brown J.~A., Wallerstein G., Geisler D., Oke J.~B., 1996, AJ, 112, 1551 
 
 \bibitem[\protect\citeauthoryear{Carraro et al.}{2007}]{carraro07} 
 Carraro G., Geisler D., Villanova S., Frinchaboy P.~M., Majewski S.~R., 
 2007, A\&A, 476, 217  


 \bibitem[\protect\citeauthoryear{Carretta et al.}{2009}]{carretta09} 
 Carretta E., Bragaglia A., Gratton R., D'Orazi V., Lucatello S., 2009, A\&A, 508, 695 

 \bibitem[\protect\citeauthoryear{Davis}{1994}]{davis94} 
 Davis L. E., 1994, A Reference Guide to the IRAF/DAOPHOT Package,
 IRAF Programming Group, NOAO, Tucson
 
 \bibitem[\protect\citeauthoryear{Dean, Warren, \& Cousins}{1978}]{dean78} 
 Dean J.~F., Warren P.~R., Cousins A.~W.~J., 1978, MNRAS, 183, 569 

 \bibitem[\protect\citeauthoryear{Dias et al.}{2002}]{dias02} 
 Dias W.~S., Alessi B.~S., Moitinho A., L{\'e}pine J.~R.~D., 2002, A\&A, 
 389, 871  

 \bibitem[\protect\citeauthoryear{Di Fabrizio et al.}{2005}]{difabrizio05} 
 Di Fabrizio L., Bragaglia A., Tosi M., Marconi G., 2005, MNRAS, 359, 966 
 
 \bibitem[\protect\citeauthoryear{Dominguez et al.}{1999}]{domin99} 
 Dominguez I., Chieffi A., Limongi M., Straniero O., 1999, ApJ, 524, 226 

 \bibitem[\protect\citeauthoryear{Durgapal, Pandey, \& Mohan}{Durgapal et al.}{2001}]{d01} 
 Durgapal A.~K., Pandey A.~K., Mohan V., 2001, A\&A, 372, 71 
  
 \bibitem[\protect\citeauthoryear{Fagotto et al.}{1994}]{fagotto94} 
 Fagotto F., Bressan A., Bertelli G., Chiosi C., 1994, A\&AS, 105, 29 

 \bibitem[\protect\citeauthoryear{Freeman \& Bland-Hawthorn}{2002}]{fbh} \
 Freeman K., Bland-Hawthorn J., 2002, ARA\&A, 40, 487 

 \bibitem[\protect\citeauthoryear{Friel}{1995}]{friel95} 
 Friel E. D., 1995, ARA\&A, 33, 381
 
 \bibitem[\protect\citeauthoryear{Friel, Jacobson, \& Pilachowski}{2005}]{friel05} 
 Friel E.~D., Jacobson H.~R., Pilachowski C.~A., 2005, AJ, 129, 2725 

 \bibitem[\protect\citeauthoryear{Friel, Jacobson, \& Pilachowski}{2010}]{friel10} 
 Friel E.~D., Jacobson H.~R., Pilachowski C.~A., 2010, AJ, 139, 1942 

 \bibitem[\protect\citeauthoryear{Friel et al.}{2002}]{friel02}  
 Friel E. D., Janes K. A., Tavarez M., Scott J., Katsanis R., Lotz J.,
 Hong  L., Miller N., 2002, AJ, 124, 2693 

 \bibitem[\protect\citeauthoryear{Frinchaboy et al.}{2004}]{frinchaboy04} 
 Frinchaboy P.~M., Majewski S.~R., Crane 
 J.~D., Reid I.~N., Rocha-Pinto H.~J., Phelps R.~L., Patterson R.~J., 
 Mu{\~n}oz R.~R., 2004, ApJ, 602, L21 

 \bibitem[\protect\citeauthoryear{Frinchaboy et al.}{2008}]{frinchaboy08} 
 Frinchaboy P.~M., Marino R.~R., et al
 2008, MNRAS, 391, 39 

 \bibitem[\protect\citeauthoryear{Frinchaboy et al.}{2006}]{frinchaboy06} 
 Frinchaboy P.~M., Mu{\~n}oz R.~R., Phelps R.~L., Majewski S.~R., Kunkel W.~E.,
 2006, AJ, 131, 922    

 \bibitem[\protect\citeauthoryear{Gratton, Sneden, \& Carretta}{2004}]{araa04} 
 Gratton R., Sneden C., Carretta E., 2004, ARA\&A, 42, 385 

 \bibitem[\protect\citeauthoryear{Ibata et al.}{2003}]{ibata03} 
 Ibata R.~A., Irwin M.~J., Lewis G.~F., Ferguson A.~M.~N., Tanvir N., 2003,  MNRAS, 340, L21 

 \bibitem[\protect\citeauthoryear{Janes \& Phelps}{1994}]{jp94} 
 Janes K. A., Phelps R. L. 1994, AJ, 108, 1773 

\bibitem[\protect\citeauthoryear{Kalirai \& Tosi}{2004}]{Kalirai04}
Kalirai, J. S., Tosi, M. 2004, MNRAS, 351, 649

 \bibitem[\protect\citeauthoryear{King et al.}{2005}]{king05} 
King I.~R., Bedin L.~R., Piotto G., Cassisi S., Anderson J., 2005, AJ, 130, 
626 

 \bibitem[\protect\citeauthoryear{Kubiak et al.}{1992}]{k92} 
 Kubiak M., Kaluzny J., Krzeminski W., Mateo M., 1992, AcA, 42, 155 
 
 \bibitem[\protect\citeauthoryear{Landolt}{1992}]{landolt92}
 Landolt A. U., 1992, AJ, 104, 340

 \bibitem[\protect\citeauthoryear{MacMinn et al.}{1994}]{macminn} 
 MacMinn D., Phelps R.~L., Janes K.~A., Friel E.~D., 1994, AJ, 107, 1806  

 \bibitem[\protect\citeauthoryear{Martin et al.}{2004}]{martin04} 
  Martin N.~F., Ibata R.~A., Bellazzini M., Irwin M.~J., Lewis G.~F., Dehnen W., 2004, MNRAS, 348, 12 

 \bibitem[\protect\citeauthoryear{Mermilliod \& Paunzen}{2003}]{merm03} 
 Mermilliod, J.~-C. \& Paunzen, E., 2003, A\&A, 410, 511  

 \bibitem[\protect\citeauthoryear{Momany et al.}{2001}]{momany} 
 Momany Y., et al., 2001, A\&A, 379, 436 
 
 \bibitem[\protect\citeauthoryear{Momany et al.}{2004}]{momanycma} 
 Momany Y., Zaggia S.~R., Bonifacio P., Piotto G., De Angeli F., Bedin L.~R., Carraro G., 2004, A\&A, 421, L29 

 \bibitem[\protect\citeauthoryear{Newberg et al.}{2002}]{newberg02} 
  Newberg H.~J., et al., 2002, ApJ, 569, 245 

 \bibitem[\protect\citeauthoryear{Ochsenbein, Bauer, \& Marcout}{2000}]{vizier} 
 Ochsenbein F., Bauer P., Marcout J., 2000, A\&AS, 143, 23 

 \bibitem[\protect\citeauthoryear{Panagia \& Tosi}{1981}]{panagiatosi} 
  Panagia N., Tosi M., 1981, A\&A, 96, 306 

 \bibitem[\protect\citeauthoryear{Phelps, Janes \& Montgomery}{Phelps et al.}{1994}]{pjm94}
 Phelps R. L., Janes K. A., Montgomery K .A., 1994, AJ 107, 1079
 
\bibitem[\protect\citeauthoryear{Phelps \& Janes}{1996}]{pj96} 
 Phelps R.~L., Janes K.~A., 1996, AJ,  111, 1604

 
 \bibitem[\protect\citeauthoryear{Ro{\v s}kar et al.}{2008}]{roskar} 
 Ro{\v s}kar R., Debattista V.~P., Quinn T.~R., Stinson G.~S., Wadsley J., 2008, ApJ, 684, L79 

 \bibitem[\protect\citeauthoryear{Rudolph et  al.}{2006}]{rudolph} 
 Rudolph A.~L., Fich M., Bell G.~R., Norsen  T., Simpson J.~P., Haas M.~R., Erickson E.~F., 2006, ApJS, 162, 346 


 \bibitem[\protect\citeauthoryear{Sch{\"o}nrich \& Binney}{2009}]{sb09} 
 Sch{\"o}nrich R., Binney J., 2009, MNRAS, 396, 203 

 
 \bibitem[\protect\citeauthoryear{Sellwood \& Binney}{2002}]{sellwood} 
 Sellwood J.~A., Binney J.~J., 2002, MNRAS, 336, 785 

 \bibitem[\protect\citeauthoryear{Sestito et al.}{2008}]{sestito08} 
  Sestito P., Bragaglia A., Randich S., Pallavicini R., Andriewsky S.M., 
  Korotin S.A., 2008, A\&A, 488, 943
 
 \bibitem[\protect\citeauthoryear{Stanghellini \& Haywood}{2010}]{sh} 
 Stanghellini L., Haywood M., 2010, ApJ, 714, 1096 

 \bibitem[\protect\citeauthoryear{Stetson}{1987}]{stetson87} 
 Stetson P. B., 1987, PASP 99, 191

 \bibitem[\protect\citeauthoryear{Tombaugh}{1938}]{tombaugh} 
 Tombaugh C., 1938, PASP, 50, 171 

 \bibitem[\protect\citeauthoryear{Tosi et al.}{1991}]{tosi91} 
 Tosi M., Greggio L., Marconi G., Focardi P., 1991, AJ, 102, 951
 
 \bibitem[\protect\citeauthoryear{Tosi, Bragaglia \& Cignoni}{Tosi et al.}{2007}]{tosi07}
 Tosi M., Bragaglia A., Cignoni M., 2007, MNRAS, 378, 730

 \bibitem[\protect\citeauthoryear{Twarog \& Anthony-Twarog}{1989}]{tat89} 
 Twarog B.~A., Anthony-Twarog B.~J., 1989, AJ, 97, 759 

 \bibitem[\protect\citeauthoryear{Twarog, Ashman, \& Anthony-Twarog}{1997}]{taat} 
 Twarog B.~A., Ashman K.~M., Anthony-Twarog B.~J., 1997, AJ, 114, 2556 

 \bibitem[\protect\citeauthoryear{Ventura et al.}{1998}]{ventura98} 
 Ventura P., Zeppieri A., Mazzitelli I., D'Antona F., 1998, A\&A, 334, 953 

 \bibitem[\protect\citeauthoryear{Villanova et al.}{2005}]{villanovabe66} 
 Villanova S., Carraro G., Bresolin F., Patat F., 2005, AJ, 130, 652  

\bibitem[\protect\citeauthoryear{Villanova et al.}{2010}]{villanovato2} 
 Villanova S., Randich S., Geisler D., Carraro G., Costa E., 2010, A\&A, 509, A102 

 \bibitem[\protect\citeauthoryear{Yong, Carney, \& Teixera de  Almeida}{Yong et
 al.}{2005}]{yong05} 
 Yong D., Carney B.~W., Teixera de Almeida M.~L., 2005, AJ, 130, 597 

 \bibitem[\protect\citeauthoryear{Wu et al.}{2009}]{wu} 
 Wu Z.-Y., Zhou X., Ma J., Du C.-H., 2009, MNRAS, 399, 2146 




\end{thebibliography}
\end{document}